# Spatially explicit accuracy assessment of deep learning-based, fine-resolution built-up land data in the United States

**Johannes H. Uhl**[1,2], **Stefan Leyk**[1,3]

[1]University of Colorado Boulder, Institute of Behavioral Science, 483 UCB, Boulder, CO-80309, USA.
[2]University of Colorado Boulder, Cooperative Institute for Research in Environmental Sciences (CIRES) 216 UCB, Boulder, CO-80309, USA.
[3]University of Colorado Boulder, Department of Geography, 260 UCB, Boulder, CO-80309, USA.

Corresponding Author: Johannes H. Uhl (Johannes.Uhl@colorado.edu)

**Abstract**: Geospatial datasets derived from remote sensing data by means of machine learning methods are often based on probabilistic outputs of abstract nature, which are difficult to translate into interpretable measures. For example, the Global Human Settlement Layer GHS-BUILT-S2 product reports the probability of the presence of built-up areas in 2018 in a global 10 m × 10 m grid. However, practitioners typically require interpretable measures such as binary surfaces indicating the presence or absence of built-up areas or estimates of sub-pixel built-up surface fractions. Herein, we assess the relationship between the built-up probability in GHS-BUILT-S2 and reference built-up surface fractions derived from a highly reliable reference database for several regions in the United States. Furthermore, we identify a binarization threshold using an agreement maximization method that creates binary built-up land data from these built-up probabilities. These binary surfaces are input to a spatially explicit, scale-sensitive accuracy assessment which includes the use of a novel, visual-analytical tool which we call focal precision-recall signature plots. Our analysis reveals that a threshold of 0.5 applied to GHS-BUILT-S2 maximizes the agreement with binarized built-up land data derived from the reference built-up area fraction. We find high levels of accuracy (i.e., county-level F-1 scores of almost 0.8 on average) in the derived built-up areas, and consistently high accuracy along the rural-urban gradient in our study area. These results reveal considerable accuracy improvements in human settlement models based on Sentinel-2 data and deep learning, in both rural and urban areas, as compared to earlier, Landsat-based versions of the Global Human Settlement Layer.

## 1. Introduction

Accurately mapping the spatial distribution and dynamics of human settlements at planetary scale is critical for monitoring and understanding global processes such as (sub)urbanization, land take, rural-urban transformations, the dynamics of the wildland-urban interface and other human-nature coupled systems. To understand, mitigate, or adapt to pressing issues related to these processes, such as biodiversity loss, overpopulation or increasing social inequality, and to ensure sustainable urban and rural development, stakeholders, planners, and researchers often use remote sensing-based land use, land cover, or settlement data as a basis for decision making. For example, the change rate of built-up area over time with respect to population change is an important indicator for sustainable urban development (Corbane et al. 2018, Ehrlich et al. 2018, Melchiorri et al. 2019, Cai et al. 2020) which may, alongside other demographic or socio-economic metrics, drive political decisions at a local scale but also at a country level or even in a global context.

Thus, the accuracy of the data underlying such decisions is critical. Global, high-resolution, and typically multitemporal datasets on built-up areas have emerged in recent years, catalyzed by the availability of long-term remote sensing archives (e.g., Landsat), the more recently launched Sentinel-1 and Sentinel-2 platforms, and by technological advances facilitating data access and processing, such as Google Earth Engine (Gorelick et al. 2017) or Deep Learning (Ball et al. 2017, Zhu et al. 2017). Such datasets include the Global Human Settlement Layer (GHSL; Pesaresi et al. 2013), Global Artificial Impervious Area (GAIA; Gong et al. 2020), Global Urban Footprint (Esch et al. 2017) and its successor, the World Settlement Footprint (WSF; Marconcini et al. 2020a) including the multi-temporal dataset WSF-Evolution (Marconcini et al. 2020b), as well as the High Resolution Settlement Layer (HRSL; Facebook Connectivity Lab & CIESIN 2016). Moreover, industry-driven efforts have sparked the availability of building footprint and road network data at a continental or nearly-planetary scale[1,2,3] (Sirko et al. 2021), complemented by Volunteered Geographic Information (i.e., OpenStreetMap, OSM). Such community-based, participatory mapping efforts can be useful for local, timely data acquisition, for example in the case of disaster response (Herfort et al. 2021). While these datasets represent considerable improvements regarding their spatial resolution as compared to older data products (e.g., Defourny et al. 2006, Balk et al. 2008), there is an urgent demand for quantifying the accuracy of these datasets, to enable informed, reflected, and uncertainty-aware data interpretation and decision making. This is particularly important as such datasets are commonly used for population disaggregation (Freire et al. 2015, Leyk et al. 2019, Palacios-Lopez et al. 2019).

To meet this demand, researchers have carried out data integration and evaluation efforts to facilitate accuracy assessments of the GHSL built-up area layers at 38 m or 30 m resolution (Blei et al. 2018, Leyk et al. 2018, Liu et al. 2020), OSM (Hecht et al. 2013, Fan et al. 2014, Brovelli & Zamboni 2018), WSF (Marconcini et al. 2020a), or cross-comparisons of several of the aforementioned datasets (Klotz et al. 2016). Such evaluation efforts also include studies focusing on specific regions (Mück et al. 2017, Sliuzas et al. 2017, Leyk et al. 2018, Liu et al. 2020, Tripathy & Balakrishnan 2021), geographic concepts (e.g., the rural-urban continuum; Leyk et al. 2018, Uhl et al. 2018, Uhl & Leyk 2022a, Uhl & Leyk 2022b), rural areas in particular (Kaim et al. 2022, Wang et al. 2022), or different landscape and settlement types (e.g., informal settlements; van den Hoek & Friedrich 2021). Related to data quality issues, the implications of discrepancies between population-based and built-up area based urban definitions have been studied (Balk et al. 2018).

A common bottleneck for unbiased and rigorous accuracy assessment of built-up land data is the limited availability of independently compiled reference data of presumably higher accuracy than the data under test (FGDC 1998). In previous work, we used an integrated dataset of cadastral parcel data and building footprint data, including construction year information, to generate multi-temporal ground truth data of built-up land at arbitrary points in time (Leyk et al. 2018, Uhl et al. 2018, Uhl & Leyk 2020, Uhl & Leyk 2022a, Uhl & Leyk 2022b). This reference dataset is the Multi-Temporal Building Footprint dataset (MTBF-33; Uhl & Leyk 2022c) which we made publicly available, and which covers 33 U.S. counties. While these efforts focused on the GHS-BUILT R2015B (38m resolution, Pesaresi et al. 2016), and the GHSL R2018A (30m resolution; Florczyk et al. 2019), it is a natural next step to apply our evaluation framework to the more recent, GHS-BUILT-S2 dataset available at 10m spatial resolution. The GHS-BUILT-S2 is one of the input layers for the most recent generation of the GHS-BUILT data suite (Schiavina et al. 2022), Thus, quantitative knowledge of its accuracy is crucial for the evaluation of the follow-up data products mapping different components of the built environment, rural-urban classes, as well as population distributions. Herein, we use our reference data employed in earlier studies, and conduct a spatially explicit accuracy assessment of the GHS-BUILT-S2 built-up land layer, in analogy to our previous work. The GHS-BUILT-S2 dataset is particularly interesting, as it reports the probability of being built-up for each grid cell, whereas previous GHS-BUILT versions provided binary built-up / not built-up labels. These "built-up probabilities" are the output of a deep learning-based classifier, and indicate the presence of built-up surface. The quantization of these values to a meaningful measure of built-up area is crucial for the unbiased usability of the GHS-BUILT-S2. This problem is a common issue in the broader context of explainable and interpretable artificial intelligence (Gunning et al. 2019, Papadakis et al. 2022), and contributes to a growing body of literature focusing on uncertainty-aware applications of deep learning in the field of remote sensing (Maxwell et al. 2021a,b). Thus, in this work, we pose the following questions:

1) How can built-up probabilities be translated into a meaningful, physical measure of built-up area?
2) How accurate is the GHS-BUILT-S2 across regions in the United States, and across the rural-urban continuum, in a localized context?

---

[1] https://sites.research.google/open-buildings/
[2] https://github.com/microsoft/GlobalMLBuildingFootprints
[3] https://github.com/microsoft/RoadDetections

3) How sensitive are the obtained, spatially explicit, thematic accuracy estimates to grid misalignment and to the chosen spatial unit?

We try to answer these questions by 1) analyzing the relationship between built-up probability and built-up area fraction, and performing an agreement analysis to find the "optimal" threshold for converting built-up probabilities in GHS-BUILT-S2 to binary layers of built-up land, 2) conducting a rigorous, spatially explicit accuracy assessment against our reference data, and 3) assessing the sensitivity of our accuracy estimates to various parameters used in the accuracy assessment. For the spatially explicit accuracy assessment conducted in step 2) we present a novel visual-analytical method to assess local accuracy variations within a given (focal) region, which we call "focal precision-recall signature plots". In the following, we describe our data and methods (Section 2), present and discuss our results (Section 3), and conclude with some final remarks (Section 4).

## 2. Data and methods

The data used in this study is the GHS-BUILT-S2 dataset, and the MTBF-33 reference dataset (Section 2.1) to carry out all analytical steps (Section 2.2).

### 2.1. Input data and preprocessing

We collected parcel data including construction year information, and building footprint data from cadastral offices for a selection of 33 counties in the United States. This work was done in 2016, when no open country-wide building footprint dataset or open harmonized parcel dataset existed. We integrated the parcel and building footprint data via spatial joins to create a set of over 6 million building footprints attributed with their construction year (Uhl & Leyk 2022c). This dataset is the MTBF-33 dataset and represents a reliable data source to create snapshots of built-up area for arbitrary points in time between 1900 and 2015. Building construction dates come from public records such as tax assessments, and building footprint data is mostly derived from LiDAR point cloud data or manual digitization. Herein, we used 30 out of the 33 counties covered by MTBF-33 as study areas, as shown in Fig. 1.

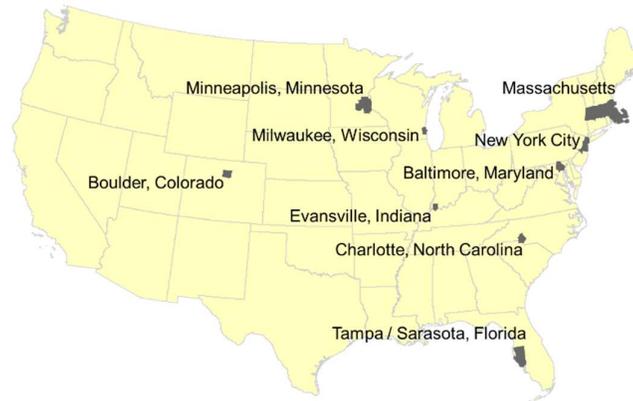

**Figure 1. Study area: The 30 counties covered by MTBF-33. These counties can be grouped into nine study regions, which are labelled in the map.**

For the counties covered by MTBF-33, we obtained the GHS-BUILT-S2 surfaces, available at 10 m spatial resolution, in Mollweide projection (Corbane et al. 2020). Among these counties are Hampden County (Massachusetts) and New York City[4], for which we show the GHS-BUILT-S2 in Fig. 2. We then rasterized the MTBF-33 building polygon data to an intermediate spatial resolution of 2 m, aligned to (and nested within) the GHS-BUILT-S2 grid, and then aggregated these 2m grid cells to the 10 m cells by calculating the proportion of 2 m cells, i.e., the fraction of reference built-up area per grid cell, reported in % (see Fig. 2 for some examples). We did this based on all buildings in MTBF-33 (regardless of their construction date), and thus, approximately representing the building stock in 2015. This rasterization process is also illustrated in Fig. 3 a) and b).

---

[4] Our New York City study area encompasses the five boroughs of Bronx, Brooklyn, Manhattan, Queens, and Staten Island.

The GHS-BUILT-S2 dataset contains built-up probabilities, which are the output of a convolutional neural network based classifier called *GHS-S2Net* (Corbane et al. 2021). GHS-BUILT-S2 is based on Sentinel-2 multispectral remote sensing data, acquired in 2018. Thus, there is a 3-year temporal offset to our reference data..

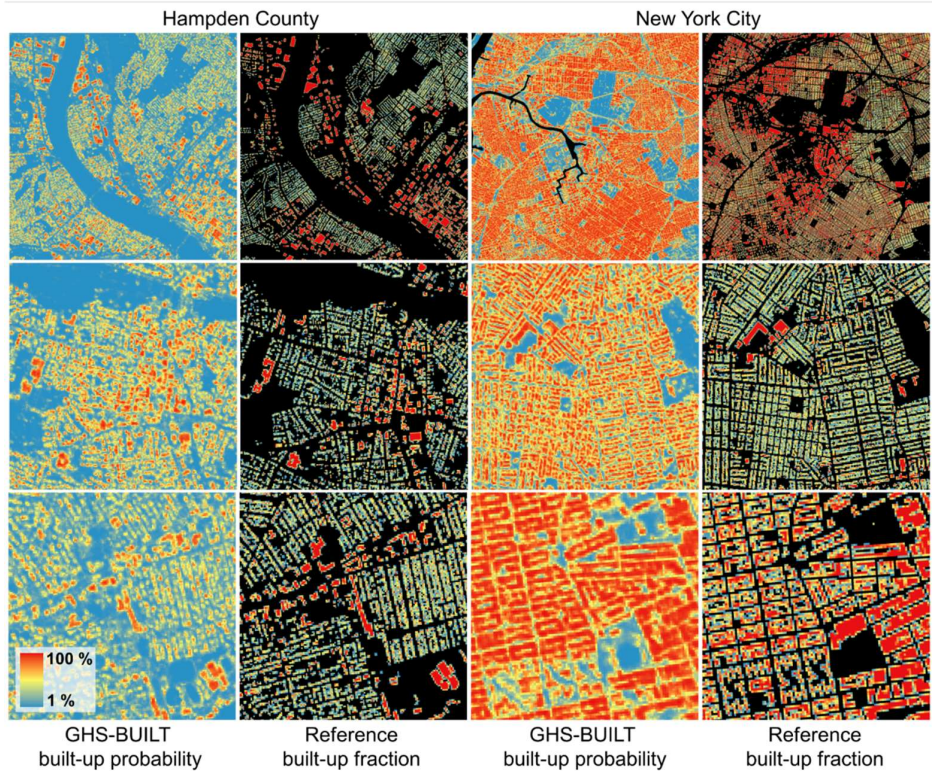

**Figure 2. Visual comparison between GHS-BUILT-S2 built-up probabilities and built-up fractions based on the reference data for parts of Hampden County (Massachusetts) and New York City.**

## 2.2. Methods

### 2.2.1. Assessing the relationship between built-up probability and built-up fraction

Based on the grid cells covered by both, GHS-BUILT-S2 and MTBF-33, we visually compared the co-occurrences of built-up probability (henceforth called BUPROB) in GHS-BUILT-S2 and the built-up fraction (henceforth called BUFRAC) derived from the reference data, on a cell-by-cell basis, for each county under study, and across all counties. We also calculated correlations between BUFRAC (see Fig. 3b) and BUPROB (see Fig. 3c) across all counties, and for each county individually. This way we were able to assess whether the relationship between BUPROB and BUFRAC is stationary across regions as an indication of the robustness of the classifier underlying the GHS-BUILT-S2 and its spatial generalization capabilities (Section 3.1). Moreover, we conducted a regression analysis between the two spatial variables, to further test the variability of their relationship across regions. Knowledge on this relationship will be important to better understand the translation of built-up probability into built-up fraction. Specifically, we used regression analysis to test whether BUFRAC can be estimated from BUPROB by fitting a function. We tested a linear model, as well as $2^{nd}$, $3^{rd}$, and $4^{th}$ order polynomial functions to allow for a more complex relationship between the two spatial variables, and analyzed regional variability of the model performances.

### 2.2.2. Generating binary built-up surface layers

For many applications of human settlement data, such as population disaggregation (e.g., Leyk et al. 2019, Palacios-Lopez 2019) or analyses of urban size and morphology (e.g., Strano et al. 2021, Uhl et al. 2021, Uhl & Leyk 2022b), practitioners require binary layers of built-up vs. not built-up areas. Thus, a second issue when working with the GHS-BUILT-S2 data product will be the thresholding / binarization of the continuous built-up probabilities to create a binary layer. In order to provide guidelines on how to identify the optimal threshold in the GHS-BUILT-S2 BUPROB values, we use a heuristic approach: We apply four cutoff values (i.e., >0%, >25%, >50%, >75%) to both the reference BUFRAC surfaces (see Fig. 3d), and the BUPROB surfaces (Fig. 3e). We then assessed the thematic agreement between the binarized BUFRAC and BUPROB surfaces, for all 16 threshold combinations, by calculating the F-1 score (van Rijsbergen 1974) for each combination, for each county, and for sub-county strata that were defined based on reference built-up area density. This is motivated by prior work, that revealed accuracy variations across the rural-urban continuum (Leyk et al. 2018, Uhl et al. 2018, Uhl et al. 2022a, Uhl et al. 2022b) creating the expectation that the relationship between BUPROB and BUFRAC could also vary between rural and urban strata. To do so, we calculated the built-up area density based on the reference built-up area fractions within 1 km × 1 km blocks. In each county, we grouped these blocks into lower-, medium- and higher-density strata, using an equal width classification scheme. Finally, for each county and each of its strata, we identified the threshold combination for which the agreement maximizes (Section 3.2).

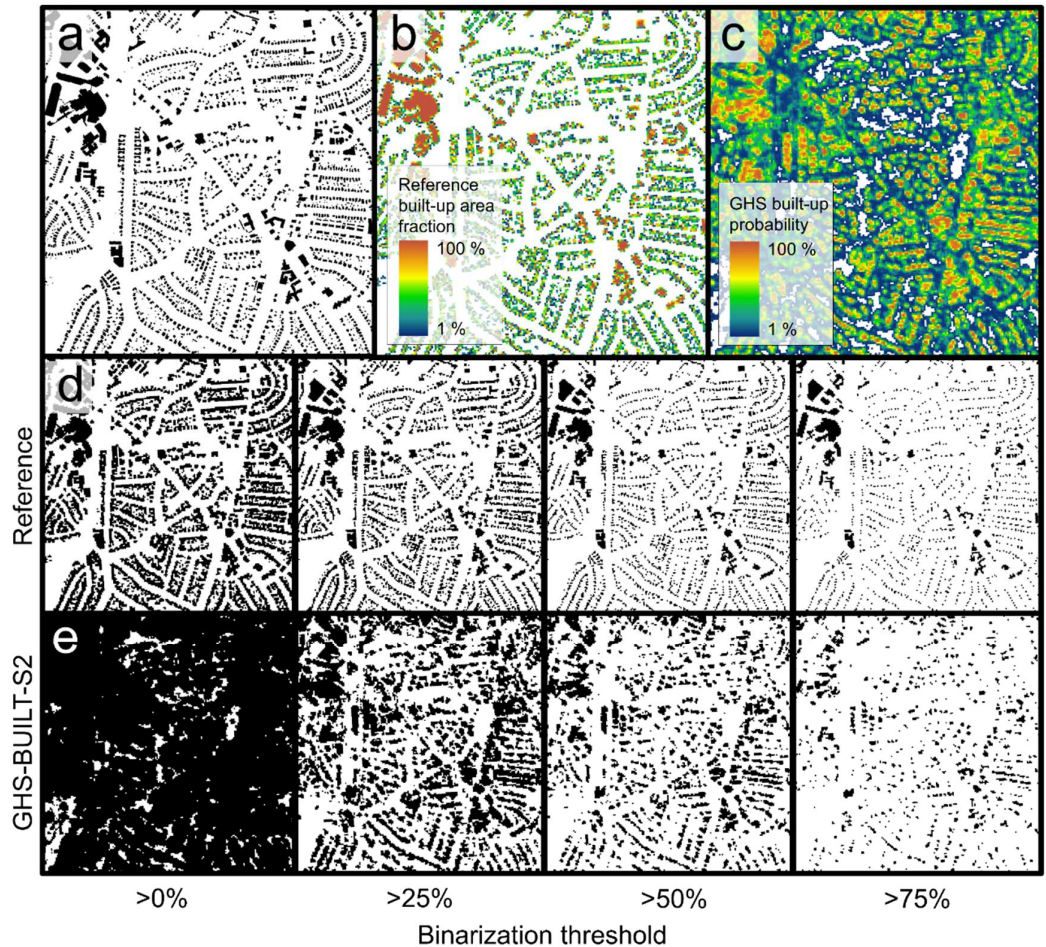

**Figure 3. Preprocessing 10m built-up probabilities and reference data to generate different binary built-up surface layers. (a) Rasterized reference building footprint data at 2m spatial resolution, derived from building footprint vector data from the MTBF-33 database, (b) reference built-up fraction at 10m spatial resolution, (c) GHS-BUILT-S2 built-up probability surface at 10m spatial resolution, (d) reference built-up fractions binarized for a range of thresholds, and (e) the GHS-BUILT-S2 built-up probabilities binarized by the same range of thresholds. All data shown for a part of the city of Charlotte (Mecklenburg County, North Carolina).**

### 2.2.3. Spatially explicit accuracy assessment

As uncertainty in geospatial data is known to be spatially non-stationary (Foody 2002, Leyk and Zimmermann 2004, Wickham et al. 2018), researchers and analysts increasingly conduct spatially explicit accuracy assessments (Foody 2007, Löw et al. 2013, Khatami et al. 2017, Waldner et al. 2017, Mitchell et al. 2018, Morales-Barquero et al. 2019), rather than reporting overall accuracy metrics which overly generalize the spatial variations of accuracy (Strahler et al. 2006). Hence, once we decided for a cutoff value to binarize both surfaces BUPROB and BUFRAC (based on the method described in Section 2.2.2), we conducted a spatially explicit, thematic accuracy assessment between the two surfaces. To do so, we employed a strategy developed in earlier work (Uhl et al. 2022a,b), which overlays a binary, gridded test surface (binarized BUPROB) on a reference surface (binarized BUFRAC) (Fig. 4a,b). This overlay creates four surfaces, each encoding the presence (1) / absence (0) of the four agreement categories per grid cell (true positves = TP, true negatives = TN, false positives = FP, false negatives = FN). Next, a user-defined kernel convolves over each of these agreement category surfaces and counts the occurrences of TP, FP, and FN within a focal window given by the extent of the kernel. Herein, we used quadratic kernels of size 1 km × 1 km, 2.5 km × 2.5 km, and 5 km × 5 km, to capture local accuracy at multiple spatial scales. The result of these convolutions are three surfaces per kernel size, holding the counts of TP, FP, and FN grid cells within the focal region around each grid cell, and thus, representing a spatialized (or localized) version of a binary confusion matrix (Fig. 4c).

Note that the TN category is disregarded here, as the true negatives (i.e., areas not built-up in test and reference data) often represent the dominant class, particularly in rural areas, and thus, we do not use agreement metrics involving the count of TN, as they tend to yield inflated or biased values if there is class imbalance (e.g., Rosenfeld & Melley 1980, Stehman & Wickham 2020). In a final step, surfaces of focal precision, recall, and F-measure are calculated on a cell-by-cell basis, resulting in spatially exhaustive, spatially explicit agreement metrics between the input surfaces (Fig. 4d), representing focal (or localized) accuracy estimates. These surfaces can then be used for further analyses.

The cell-by-cell sum of focal TP and FN counts yields the total number of built-up reference grid cells per focal window. Likewise, the cell-by-cell sum of focal TP and FP counts yields the total number of built-up grid cells in the test data. These counts represent the quantity of built-up area per focal window, which we call "built-up quantity". By comparing these counts derived from the reference and the test data, we can measure the quantity agreement of built-up area per focal window, while relaxing the constraint of positional alignment, as measured by the cell-by-cell agreement metrics precision, recall, and F-1 score.

Moreover, we use these counts to calculate the reference built-up surface density (henceforth called "built-up density") per focal window to define density-based strata. We then calculate accuracy metrics within these strata and analyze accuracy trajectories across the rural-urban continuum (i.e., from rural low-density settlements to urban high-density settlements) (Section 3.3).

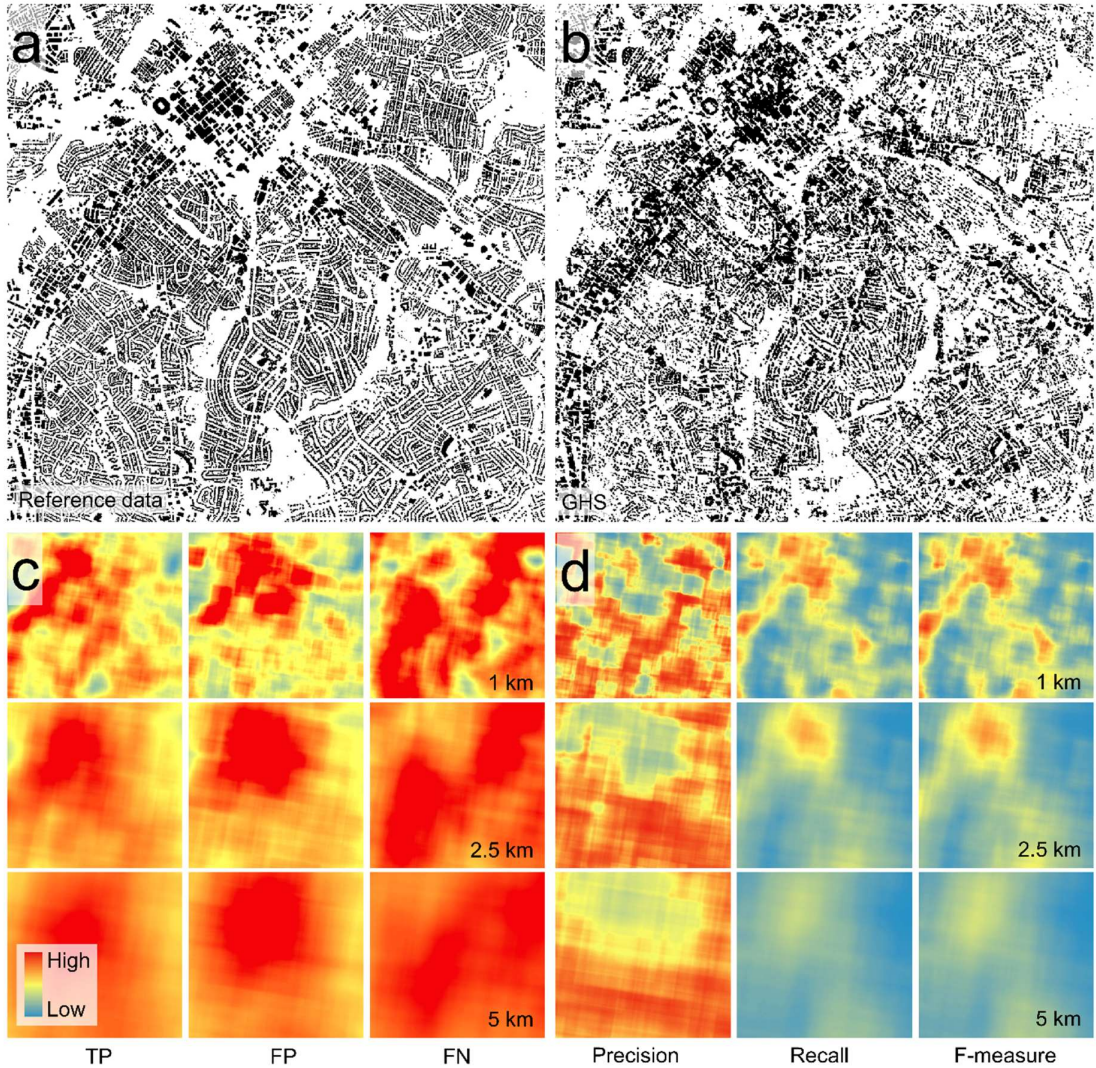

**Figure 4. Input and output surfaces of the spatially explicit accuracy assessment. (a) binarized GHS built-up probability (>0.5) and (b) binarized reference built-up fractions (>0), Also shown are (c) derived localized confusion matrix element surfaces and (d) focal accuracy surfaces. All data shown for a part of the city of Charlotte (Mecklenburg County, North Carolina).**

Herein, we used the created agreeement surfaces to perform the following analyses: We calculated the overall agreement between the binary surfaces 1) per county, 2) within three (equal width) reference built-up density-based strata per county, loosely related to rural, peri-urban, and urban areas, and 3) we visually assessed the interactions between localized commission and omission errors, by creating scatterplots of focal precision and focal recall, per county. This is motivated by the findings of previous accuracy assessments, where we found low precision, but high recall in urban areas, as well as low precision and low recall in rural areas, mostly caused by the false detection of roads as built-up areas (e.g., Leyk et al. 2018). Thus, such "signature plots" of focal precision vs focal recall provide a visual-analytical way to assess the overall levels (and distributions) of commission and omission errors, and to detect interactions between them (Section 3.3).

**2.2.4. Localized built-up area regression analysis**

While we assessed the thematic agreement between the gridded surfaces in Section 2.2.3, we are aware that positional uncertainty in the source datasets may cause misalignment between the gridded reference and test surfaces, which may bias the outcomes of our thematic accuracy assessment as described in Section 2.2.3 (e.g., Congalton 2008). Hence, we relaxed the constraint of cell-by-cell alignment between test and reference data, and used the focal built-up area in both datasets to assess their quantity agreement, disregarding their spatial overlap. We conducted linear regression analyses using the focal built-up quantities in test and reference data, per county, and across all counties, to test the regional variability of the quantity agreement (Section 3.4). This analysis is also motivated by a range of applications of human settlement data that do not require an analytical unit of 10 m grid cells, but where coarser spatial resolutions are sufficient (e.g., urban shape analysis using landscape metrics, population disaggregation, etc.).

**2.2.5. Sensitivity to positional uncertainty and assessment support**

Related to the aforementioned potential misalignment between the gridded test and reference surfaces, we analyzed the sensitivity of the agreement measures to the potential positional uncertainty in the data. As our reference surfaces are based on cadastral data, there is some positional uncertainty associated with the building footprint polygons, that propagates into our gridded surfaces (Congalton 2008). Moreover, the Sentinel-2A data underlying the GHS-BUILT-S2 dataset may be affected by positional uncertainty due to image registration and other image processing steps.

To test the sensitivity of our focal accuracy metrics to such potential misalignment, we implemented two sensitivity analyses: 1) For a selection of three counties (i.e., New York City, Hampden County, Massachusetts, and Boulder County, Colorado) we systematically shifted the reference surface by 1 and 2 cells in both x- and y-direction, mimicking offsets between the data of up the 20 m in each direction. We then recalculated the focal accuracy metrics for each combination of shifts, and assessed their variation, on average, as well as spatially explicit (Section 3.5.1).

Moreover, we recalculated our focal accuracy assessments based on aggregated grid cell blocks. For example, if a built-up grid cell at native 10-m resolution does not spatially coincide with a built-up reference cell, but has a built-up reference cell within its 3x3 cell neighborhood, the assigned agreement category would still be TP (true positive). Such an aggregation of the analytical unit used for map comparison is a commonly employed technique to relax the requirement of alignment at the native resolution and to account for potential positional discrepancies between datasets which may bias the thematic agreement measures (Congalton 2007, Gu & Congalton 2020, Gu & Congalton 2021, Marconcini et al. 2020a). We did such a block-based accuracy assessment for all counties using 3×3 cell blocks, and 5×5 cell blocks (Section 3.5.2). The data processing and analyses conducted herein are illustrated in Fig. 5.

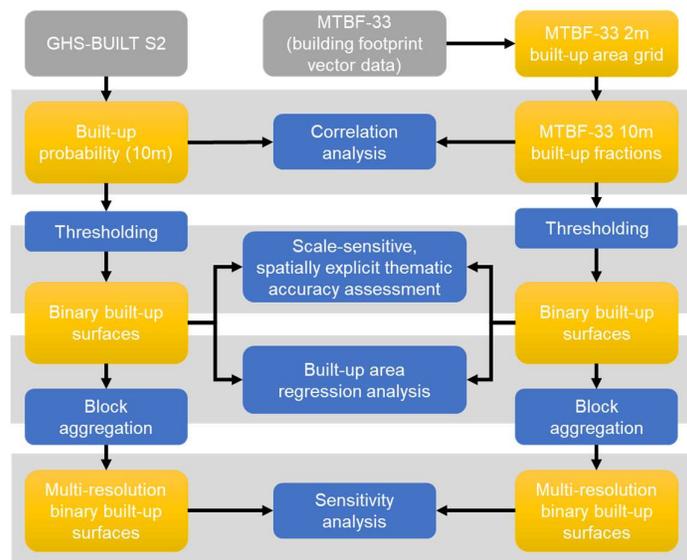

**Figure 5. Data processing and analysis flow diagram.**

# 3. Results and discussion

### 3.1. The relationship between built-up probability and built-up fraction

We binned the BUFRAC and BUPROB values for each county into bins of 4% and 1%, respectively, and visualized the bivariate histograms of the joint BUFRAC / BUPROB distribution. Moreover, we calculated the median BUPROB per BUFRAC bin. We observe generally a positive association between BUFRAC and BUPROB when using the data distribution over all counties (Fig. 6a), but we observe considerable variation in the bivariate histograms and median lines for individual counties (Fig. 6b,c, see Appendix Fig. A1 for all 30 counties under study). For example, we observe higher median BUPROB per BUPRAC bin in New York City (Fig. 6c) than in Hampden County (Fig. 6b). This effect can also be seen in the examples shown in Fig. 2, illustrating how large and densely built structures in New York City cause high levels of built-up probability, as opposed to smaller, less densely arranged buildings in Hampden County. This could also indicate that the classifier used to create the GHS-BUILT-S2 surface is more confident in detecting high-density structures as built-up, possibly also due to spillover effects caused by highly impervious surfaces in the vicinity of the buildings. Also, New York City may have been used as a training site for the GHS-BUILT-S2 classifier training phase, and the lower probabilities (which can be interpreted as lower levels of confidence) are the effect of weaker inference performance when generalizing on out-of-distribution data.

The county-level median BUPROB lines (Fig. 6d) indicate that even grid cells of very low BUFRAC have median BUPROB values of around 25% in Hampden County, and up to 60% in New York City. Thus, these two counties represent extreme cases among the counties under test. This is in line with the previous observations, and likely to be the result of spillover effects of impervious surface surrounding the buildings, as well as mixed-pixel effects in general, and the intensity of this effect is driven by the built-up density and the level of impervious surface. Another interesting observation is the drop of median BUPROB for grid cells of close to 100% built-up fraction in some counties. These are likely large buildings, possibly with a roof material that may occur more rarely in the training data, and this sparseness may cause lower levels of classification confidence.

Here, it is noteworthy that only grid cells have been taken into account where both BUFRAC and BUPROB are >0, to avoid omission or commission errors to interfere in the analysis of the relationship between BUFRAC and BUPROB.

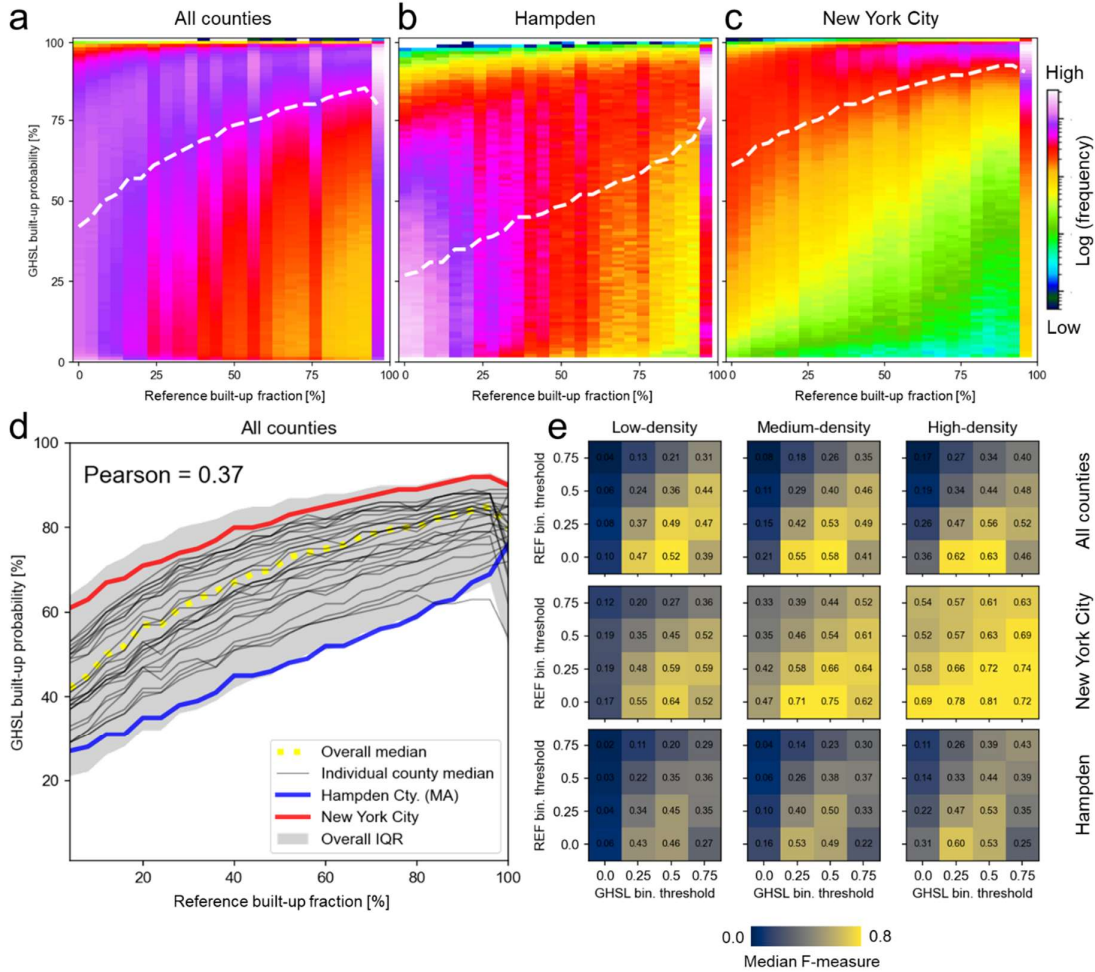

**Figure 6. Statistical relationship between built-up probability and built-up fraction for 30 U.S. counties; agreement assessment for different binarization thresholds.**

Despite the difference in the intercepts of the median BUPROB lines, the correlation coefficients between BUFRAC and BUPROB are relatively stable across counties, ranging from 0.25 to 0.4 for the majority of counties (see Appendix Table A1). The overall Pearson's correlation coefficient between BUFRAC and BUPROB is 0.37, further confirming the generally observed (and expected) positive association of BUFRAC with BUPROB.

Can users infer the built-up fraction from the built-up probability at the cell-level? To address this, we used a curve fitting approach to estimate BUFRAC as a function of BUPROB. As can be seen in Fig. 6 a-c, a given BUFRAC value corresponds to a wide range of BUPROB values, and thus, this relationship is ambiguous. The patterns observed in Fig. 6 a-c suggest a non-linear relationship between BUFRAC and BUPROB, and thus, we fitted a range of polynomial functions of degree one to four using ordinary least squares. As suspected, the predictive power of BUPROB for BUFRAC is low, with $R^2$ values rarely exceeding 0.2, and RMSE values of around 30% when estimating BUFRAC in % grid cell area (Appendix Table A2). These poor results indicate that built-up area fraction at 10 m resolution cannot be inferred reliably from the built-up probabilities provided by the classifier underlying the GHS-BUILT-S2.

### 3.2. Agreement between binary built-up / not built-up surfaces

When binarizing both, the continuous BUFRAC and BUPROB surfaces based on the cut-off values discussed in Section 2.2.2, we observe that the thematic agreement between these binarized surfaces is highest when using a threshold of >50% for the built-up probabilities, and a threshold of >0% for the reference built-up fraction surfaces. This observation is remarkably consistent across the three density-based strata (loosely related to rural, peri-urban, and urban areas), over all counties, for Hampden County and New York City (Fig. 6e) and for most of the other counties under study (Table 1). In practice, this implies that by using a cutoff value of 50% applied to the GHS-BUILT-S2 surface, the resulting binary layer can be interpreted as a built-up area presence-absence surface, and this surface exhibits relatively high agreement with a binary surface that maps any grid cell containing a built-up area fraction >0 as "built-up". Specifically, the F-1 scores range from around 0.5 in the low-density strata to 0.65 or higher in the high-density strata. While these values were obtained at a native resolution of 10 m, the F-1 scores derived from 3×3 cell blocks (corresponding to 30 m × 30 m) increase in many counties to around 0.8 or higher, measured across all three strata (Table 1). These F-1 scores obtained at 3×3 cell blocks are comparable to results of accuracy assessments carried out in earlier work, using the Landsat-based GHS-BUILT-R2018A. However, these earlier experiments resulted in considerably lower F-1 scores across rural-urban gradients (cf. Leyk et al. 2018). This indicates that a remarkable jump in accuracy can be expected when using the GHS-BUILT-S2 as a basis for built-up area mapping, at least for the areas under study, likely for the US, and possibly elsewhere as well. Moreover, these findings are in line with accuracy estimates reported in Corbane et al. 2021. Here, it should be noted that while we found high levels of consistency of the agreement maximization threshold for our study sites in the U.S., these cut-off values are likely to be different in Europe or Asia, as indicated in Hafner et al. (2022).

**Table 1. Agreement-maximization threshold for Reference data and GHSL, across different strata of urbanness, and corresponding F-1 scores per stratum, global (i.e., per county) at full resolution, and global within 3x3 cell blocks.**

| Stratum | Agreement maximization threshold | | | | | | F-1 score | | | | |
|---|---|---|---|---|---|---|---|---|---|---|---|
| | Low-dens. | | Medium-dens. | | High-dens. | | Low-dens. | Medium-dens. | High-dens. | overall 1x1 | overall 3x3 |
| County | Ref. | GHSL | Ref. | GHSL | Ref. | GHSL | | | | | |
| Anoka County | 0% | 50% | 0% | 50% | 0% | 50% | 0.612 | 0.659 | 0.667 | 0.581 | 0.715 |
| Baltimore County | 0% | 50% | 0% | 50% | 0% | 50% | 0.507 | 0.575 | 0.669 | 0.671 | 0.804 |
| Barnstable County | 25% | 50% | 0% | 50% | 0% | 50% | 0.473 | 0.587 | 0.617 | 0.613 | 0.803 |
| Berkshire County | 0% | 50% | 0% | 25% | 0% | 25% | 0.533 | 0.536 | 0.590 | 0.565 | 0.706 |
| Boulder County | 0% | 25% | 0% | 25% | 0% | 25% | 0.462 | 0.455 | 0.554 | 0.436 | 0.571 |
| Bristol County | 0% | 50% | 0% | 50% | 0% | 50% | 0.538 | 0.600 | 0.640 | 0.660 | 0.816 |
| Carver County | 0% | 50% | 0% | 50% | 0% | 50% | 0.632 | 0.667 | 0.681 | 0.588 | 0.704 |
| Dakota County | 0% | 50% | 0% | 50% | 0% | 50% | 0.533 | 0.601 | 0.669 | 0.668 | 0.811 |
| Dukes County | 0% | 50% | 0% | 50% | 0% | 50% | 0.471 | 0.522 | 0.576 | 0.581 | 0.748 |
| Essex County | 0% | 25% | 0% | 25% | 0% | 25% | 0.544 | 0.596 | 0.637 | 0.624 | 0.821 |
| Franklin County | 0% | 50% | 0% | 50% | 0% | 50% | 0.500 | 0.525 | 0.562 | 0.579 | 0.705 |
| Hampden County | 0% | 50% | 0% | 25% | 0% | 25% | 0.457 | 0.531 | 0.600 | 0.516 | 0.698 |
| Hampshire County | 0% | 50% | 0% | 50% | 0% | 25% | 0.500 | 0.519 | 0.581 | 0.552 | 0.702 |
| Hennepin County | 0% | 50% | 0% | 50% | 0% | 50% | 0.632 | 0.645 | 0.677 | 0.625 | 0.777 |
| Hillsborough County | 0% | 50% | 0% | 50% | 0% | 50% | 0.383 | 0.472 | 0.643 | 0.624 | 0.785 |
| Manatee County | 0% | 50% | 0% | 50% | 0% | 50% | 0.500 | 0.556 | 0.695 | 0.688 | 0.828 |
| Mecklenburg County | 0% | 25% | 0% | 50% | 0% | 25% | 0.534 | 0.619 | 0.659 | 0.615 | 0.753 |
| Middlesex County | 0% | 50% | 0% | 25% | 0% | 25% | 0.518 | 0.581 | 0.638 | 0.410 | 0.535 |
| Milwaukee County | 25% | 50% | 0% | 25% | 0% | 25% | 0.446 | 0.590 | 0.666 | 0.408 | 0.584 |
| Monmouth County | 0% | 50% | 0% | 50% | 0% | 50% | 0.519 | 0.625 | 0.698 | 0.507 | 0.609 |
| Nantucket County | 0% | 50% | 0% | 50% | 0% | 50% | 0.441 | 0.532 | 0.568 | 0.585 | 0.765 |
| Norfolk County | 0% | 25% | 0% | 25% | 0% | 25% | 0.554 | 0.601 | 0.644 | 0.642 | 0.824 |
| New York City | 0% | 50% | 0% | 50% | 0% | 50% | 0.637 | 0.752 | 0.810 | 0.797 | 0.929 |
| Plymouth County | 0% | 50% | 0% | 50% | 0% | 50% | 0.494 | 0.570 | 0.618 | 0.621 | 0.787 |
| Ramsey County | 0% | 50% | 0% | 50% | 0% | 50% | 0.558 | 0.636 | 0.664 | 0.636 | 0.810 |
| Sarasota County | 0% | 50% | 0% | 50% | 0% | 50% | 0.546 | 0.607 | 0.692 | 0.685 | 0.844 |
| Suffolk County | 0% | 50% | 0% | 50% | 0% | 25% | 0.591 | 0.688 | 0.743 | 0.731 | 0.906 |
| Vanderburgh County | 0% | 50% | 0% | 25% | 0% | 25% | 0.617 | 0.598 | 0.658 | 0.655 | 0.804 |
| Washington County | 25% | 75% | 0% | 50% | 0% | 50% | 0.556 | 0.634 | 0.659 | 0.604 | 0.725 |
| Worcester County | 0% | 25% | 0% | 25% | 0% | 25% | 0.482 | 0.543 | 0.596 | 0.346 | 0.473 |

### 3.3. Spatially explicit accuracy assessment

Based on the localized confusion matrices calculated in focal windows of 1 km × 1 km, we created focal precision-recall plots for each county (Fig. 7). The location, spread, and shape of the visualized point clouds provide rich information about the overall levels of commission and omission error, their variation within each county, and their interaction (Fig. 7). Thus, these plots represent a visual-analytical method to assess different aspects of map classification accuracy or agreement of binary surfaces within a given region, and to compare across regions. The additional color-coding of the data points indicating the reference built-up surface density within each focal window (scaled across all regions) enables the localization of specific commission-omission error combinations along the rural-urban continuum.

For example, New York City and Suffolk County (i.e., the city of Boston, Massachusetts) are among the counties of very high density, exhibiting high levels of completeness (i.e., recall) and correctness (i.e., precision) across their spatial extents. Surprisingly, Milwaukee County (i.e., the city of Milwaukee, Wisconsin) has similar levels of built-up surface density, but shows lower levels of precision, and even more pronounced, of recall. Mecklenburg County (i.e., the city of Charlotte, North Carolina) shows high levels of precision, but also higher levels of omission errors. Other point cloud shapes such as for Hillsborough or Sarasota County, and to a lesser degree, for Manatee County (all located in the Greater Tampa region, Florida), we see a more linear relationship of precision and recall, indicating that in places where the binarized GHS-BUILT-S2 surface is more correct, it is also more complete. This could also indicate that data from Florida was used for training the classifier underlying GHS-BUILT-S2, or the detection of settlements in that region is more straightforward than in others, possibly due to vegetation and other factors.

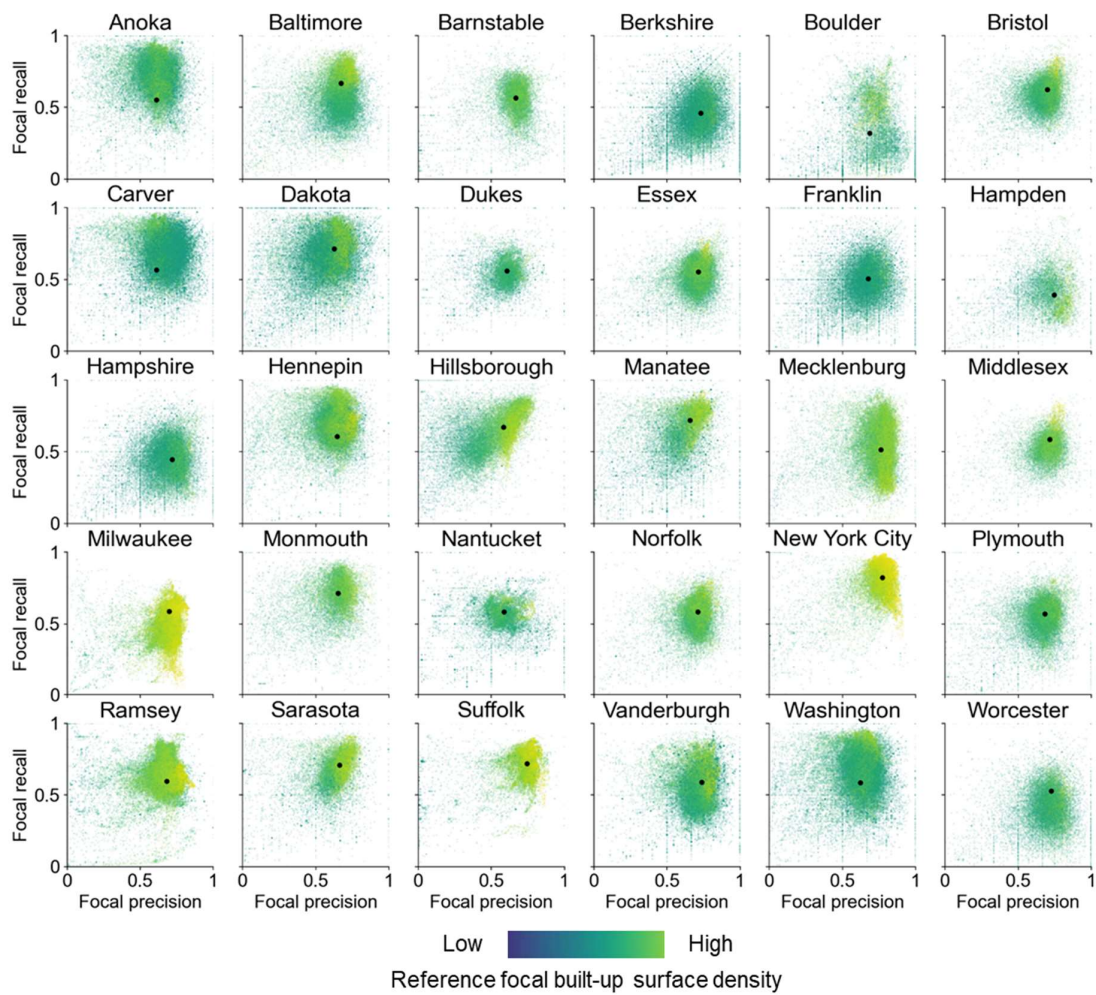

**Figure 7. Focal precision-recall plots for all 30 counties, for a focal window size of 1 km. Coordinates of the black dots show the precision and recall, respectively, calculated over the total county extents.**

These plots also reveal other interesting cases, such as Boulder County (Colorado), which seems to be divided into a high precision and high recall region (likely the city of Boulder itself), and places of high precision but low recall (possibly the scattered, rural settlement in the Mountains or the Plains). Counties with data points spread towards the left part of the graphs (i.e., low levels of precision) could be affected by recent growth in built-up area (during 2015 to 2018) that is not contained in the MTBF-33 reference data (dated to 2015 or earlier in some counties), but correctly measured in the GHS-BUILT-S2 (reflecting the state of built-up areas in 2018). Thus, these locations could contain higher levels of false positives, induced by missingness in the reference data. Finally, the black dots illustrating the overall precision-recall pairs calculated across all grid cells per county once more illustrate the need for spatially explicit accuracy assessments, as overly generalized accuracy metrics often fail to take into account the spatial non-stationarity of the uncertainty in geospatial data (e.g., Leyk and Zimmermann 2004, Strahler et al. 2005, Foody 2007, Wickham et al. 2018).

We also used our focal accuracy estimates to assess the accuracy variations along the rural-urban continuum. While there are multiple ways to model the gradient from rural to urban areas (e.g., Waldorf & Kim 2015), we used the built-up surface density derived from the reference data, as it is enumerated consistently to our focal accuracy estimates (see Section 2.2.3). We did this visually-analytically by transforming the reference built-up densities into the range [0,1] and plotting them against the F-1 score computed for each focal window (Fig. 8). These plots indicate that low accuracy (as measured by the F-1 score) almost exclusively occur in low-density regions. F-1 levels then increase steadily at low slope towards high-density regions. In some cases, F-1 scores slightly drop towards the high-density regions, likely due to mixed-pixel effects and spillover effects of impervious surfaces, in highly built-up areas, resulting in higher levels of commission errors. Generally, these trends are very encouraging, as compared to similar assessments of earlier versions of the GHS-BUILT, where the trends across the rural-urban continuum were much steeper, indicating an improved mapping of rural and peri-urban settlements in the GHS-BUILT-S2 dataset and less difference in data quality between rural and urban settings.

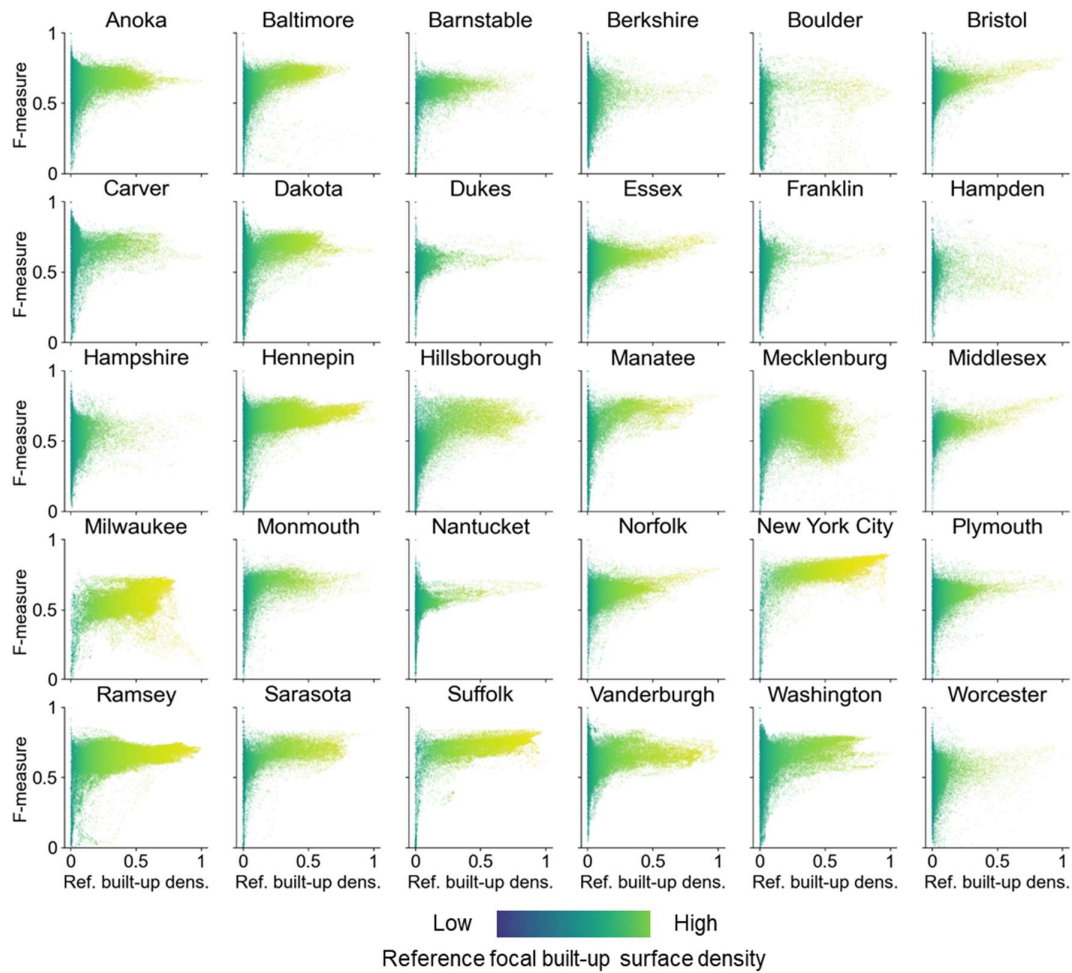

**Figure 8. Focal accuracy estimates along the rural-urban continuum: Relationship between F-measure and reference built-up density per county, for a focal support of 1 km.**

### 3.4. Regression analysis

While the assessment presented in Section 3.3 focused on thematic agreement of built-up vs. not built-up grid cells, we also assessed the quantity agreement of built-up surface within our focal windows, by means of regression analyses, relaxing the constraint of spatial coincidence of built-up grid cells. Using a linear regression model to estimate reference built-up quantity based on the GHS-derived built-up quantity (measured within the 1 km × 1 km focal windows, ranging from 0 to 1) , we find a highly linear relationship with an $R^2$ of 0.93 across all counties, a slope of 0.87 and an intercept of 0.20. These regression models perform similarly at the county level, with $R^2$ values of > 0.9 in most counties. The slope values, however, exhibit some variations across counties, with a minimum value of 0.44 in Hampden County. As we have seen in Fig. 6, Hampden County is also the county with the lowest built-up probability, on average.

Thus, the observed differences in the relationship between built-up quantities are likely an effect of lower levels of classification confidence in the GHS-BUILT-S2 in some regions. The scatterplots of focal built-up quantity from the GHS and the reference data confirm these observations (Fig. 9). In some counties we observe superposed effects of a linear relationship for low and medium density areas, and a superlinear relationship towards high-density areas, where the built-up quantity in GHS-BUILT-S2 exceeds the reference built-up quantity (e.g., Bristol, Essex, Middlesex counties). The opposite trend can be observed for Suffolk County (i.e., the city of Boston), where in high-density areas the relationship becomes sublinear.

**Table 2. Built-up quantity regression analysis results per county.**

| County | Slope | Intercept | $R^2$ | County | Slope | Intercept | $R^2$ |
|---|---|---|---|---|---|---|---|
| Anoka County | 1.102 | 0.456 | 0.868 | Manatee County | 1.046 | 0.434 | 0.932 |
| Baltimore County | 1.020 | -0.155 | 0.946 | Mecklenburg County | 0.549 | 1.913 | 0.965 |
| Barnstable County | 0.847 | 0.007 | 0.953 | Middlesex County | 0.873 | -0.798 | 0.686 |
| Berkshire County | 0.602 | 0.064 | 0.928 | Milwaukee County | 0.654 | 1.462 | 0.958 |
| Boulder County | 0.655 | -0.104 | 0.928 | Monmouth County | 0.994 | 0.939 | 0.697 |
| Bristol County | 0.936 | -0.314 | 0.875 | Nantucket County | 0.904 | 0.412 | 0.930 |
| Carver County | 1.177 | 0.047 | 0.966 | Norfolk County | 0.853 | -0.322 | 0.958 |
| Dakota County | 1.078 | 0.410 | 0.914 | New York City | 1.054 | 0.601 | 0.951 |
| Dukes County | 0.864 | 0.184 | 0.948 | Plymouth County | 0.823 | 0.065 | 0.900 |
| Essex County | 0.810 | -0.335 | 0.964 | Ramsey County | 0.811 | 2.616 | 0.946 |
| Franklin County | 0.702 | 0.075 | 0.967 | Sarasota County | 1.018 | 0.689 | 0.918 |
| Hampden County | 0.441 | 0.395 | 0.958 | Suffolk County | 0.947 | 0.818 | 0.963 |
| Hampshire County | 0.570 | 0.161 | 0.858 | Vanderburgh County | 0.760 | 0.342 | 0.938 |
| Hennepin County | 0.983 | 1.470 | 0.908 | Washington County | 1.115 | 0.131 | 0.947 |
| Hillsborough County | 1.011 | 1.454 | 0.938 | Worcester County | 0.598 | 0.010 | 0.928 |

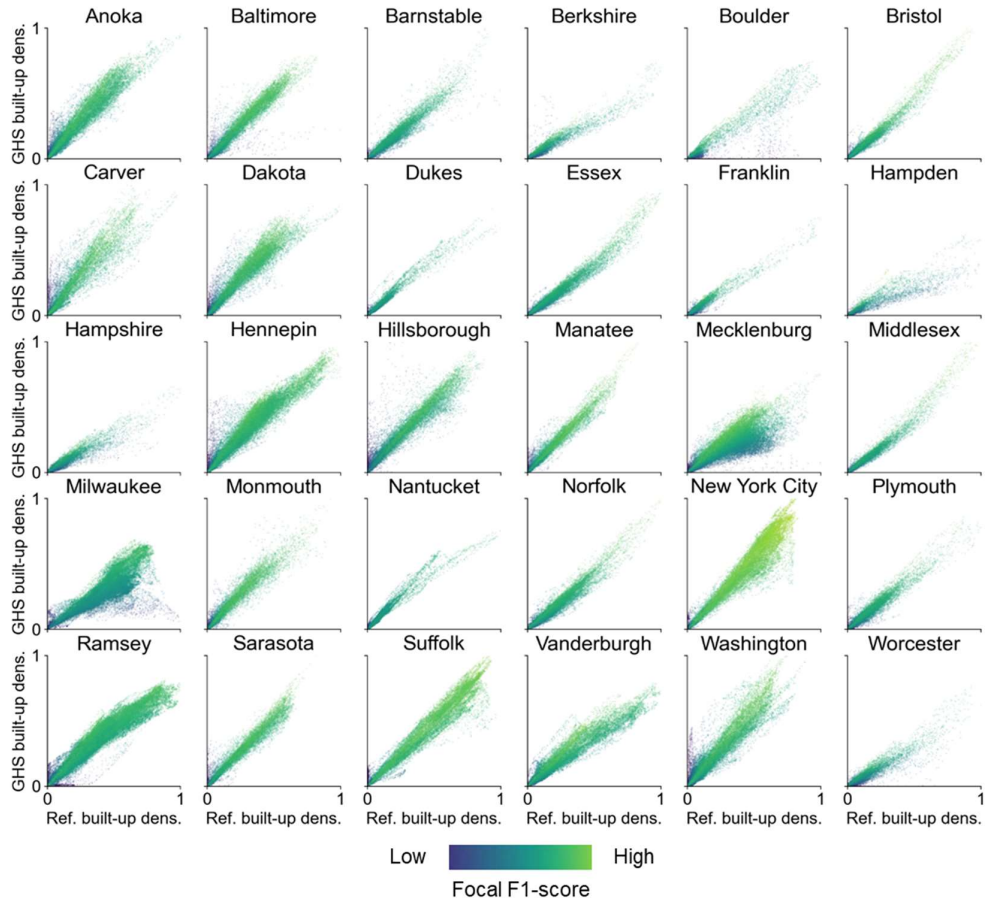

**Figure 9. Scatterplots of reference and GHS built-up quantity per county for a focal support level of 1 km.**

### 3.5. Sensitivity analysis

As noted earlier, the outcomes of thematic accuracy assessments based on gridded data may be biased by positional uncertainty in the underlying spatial data, that can result in random or even systematic misalignment of grid cells. This issue is addressed in Section 3.5.1. Moreover, the analytical unit at which the assessment is conducted, may also affect the results, due to the same misalignment but also general aggregation issues, subject to the Modifiable Areal Unit Problem (MAUP; Openshaw and Taylor 1979, Flowerdew et al. 2001, Goodchild 2022). Finally, the level of spatial support (i.e., the focal window size used to calculate localized accuracy estimates) may affect the results, which is another manifestation of the MAUP. Hence, we systematically varied both the analytical unit and the spatial support to assess the effects on our focal precision and recall estimates (Section 3.5.2).

#### 3.5.1. Sensitivity to spatial offsets

We systematically shifted the reference grid by 1 and 2 grid cells in each direction, and recomputed the focal precision and recall values for each combination of grid shifts in x-and y-direction. To keep the computational effort to a feasible level, we only conducted this analysis for three counties (i.e., Boulder County, Hampden County, and New York City). Fig. 10 reports the average focal precision and recall for each shift combination. We observe a sharp drop of average precision and recall, even if grids are shifted by 10 m only. This effect is most nuanced in Boulder County, where this drop from 0.64 to around 0.45 for precision (Fig. 10a), and from 0.2 to 0.13 for recall (Fig. 10d), corresponds to relative decreases in average agreement (i.e., approximately 27% for precision, and approximately 45% for recall, respectively). This effect is least dominant in the highly urban study area of New York City (Fig. 10 c,f; around 10% relative drop when shifting grids by 10 m), which can be explained by the spillover and mixed-pixel effects occurring in highly impervious areas. This stark contrast between high-density settings and rural counties (i.e., Boulder County is largely characterized by scattered, rural settlements in the Mountains and the Plains) emphasizes the importance of the positional accuracy of spatial data as a prerequisite for unbiased thematic accuracy assessments.

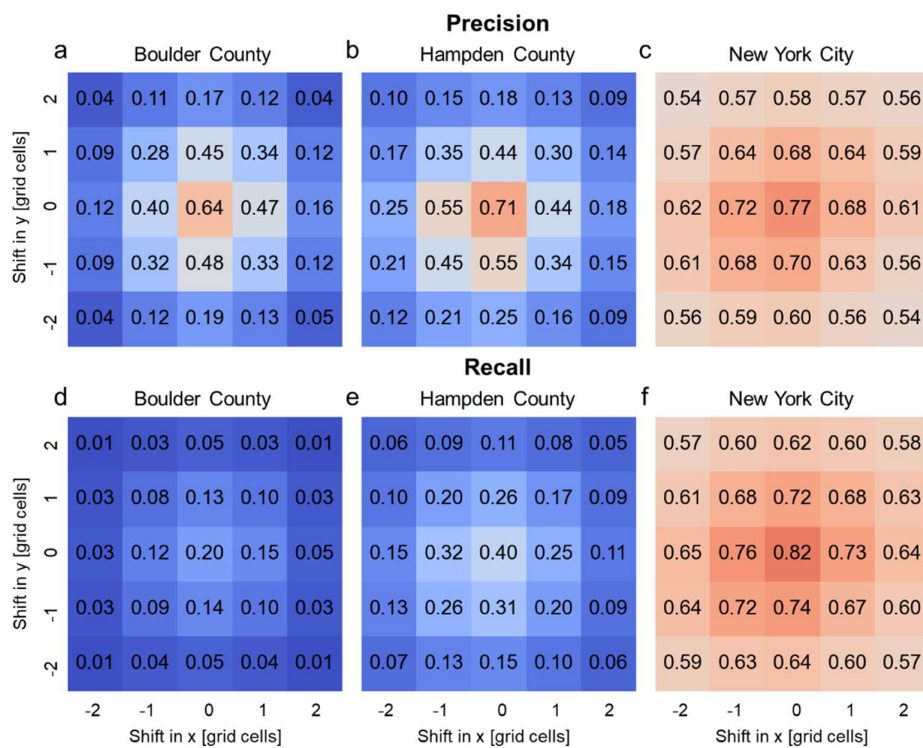

**Figure 10. Sensitivity of focal accuracy estimates to grid shifts.** Average focal precision in (a) Boulder County, (b) Hampden County, and (c) New York City, and (d-f) corresponding plots for average focal recall, based on 24 combinations of shifts in x and y direction, applied to the reference data.

While these shifts appear to affect the overall level of accuracy estimates as a function of the settlement density in a given region, their effect on the accuracy trajectories is less nuanced (Fig. 11a): While the trend lines are shifted along the y-axis (as a result of lower agreement), their slopes along the rural-urban continuum remain largely unaffected, except for areas of extremely low settlement density in Hampden and Boulder counties. Finally, we visualized the range of F-1 scores at the grid cell level, for each of the 25 grid shift scenarios (Fig. 11b). These maps reveal further detail about our previously found relationship between settlement density and accuracy sensitivity: Even within our three test counties, the sensitivity of accuracy estimates to positional offsets varies, and these variations exhibit an inverse trend to settlement density (Fig. 11c), with lowest F-1 score dispersion levels found in high-density areas within a county.

These results illustrate that accuracy estimates are differently affected by positional uncertainty in the underlying spatial data. As we can generally assume that geospatial data quality is higher in high-density urban areas than in rural regions, which may also affect our reference data (e.g., terrain variations, occlusions from vegetation, less frequent data update cycles in rural areas), it is reasonable to assume that our accuracy estimates in rural areas may be negatively biased by such positional inaccuracies, and that the "true" accuracy of the GHS-BUILT-S2 data in rural areas is even higher than reported herein.

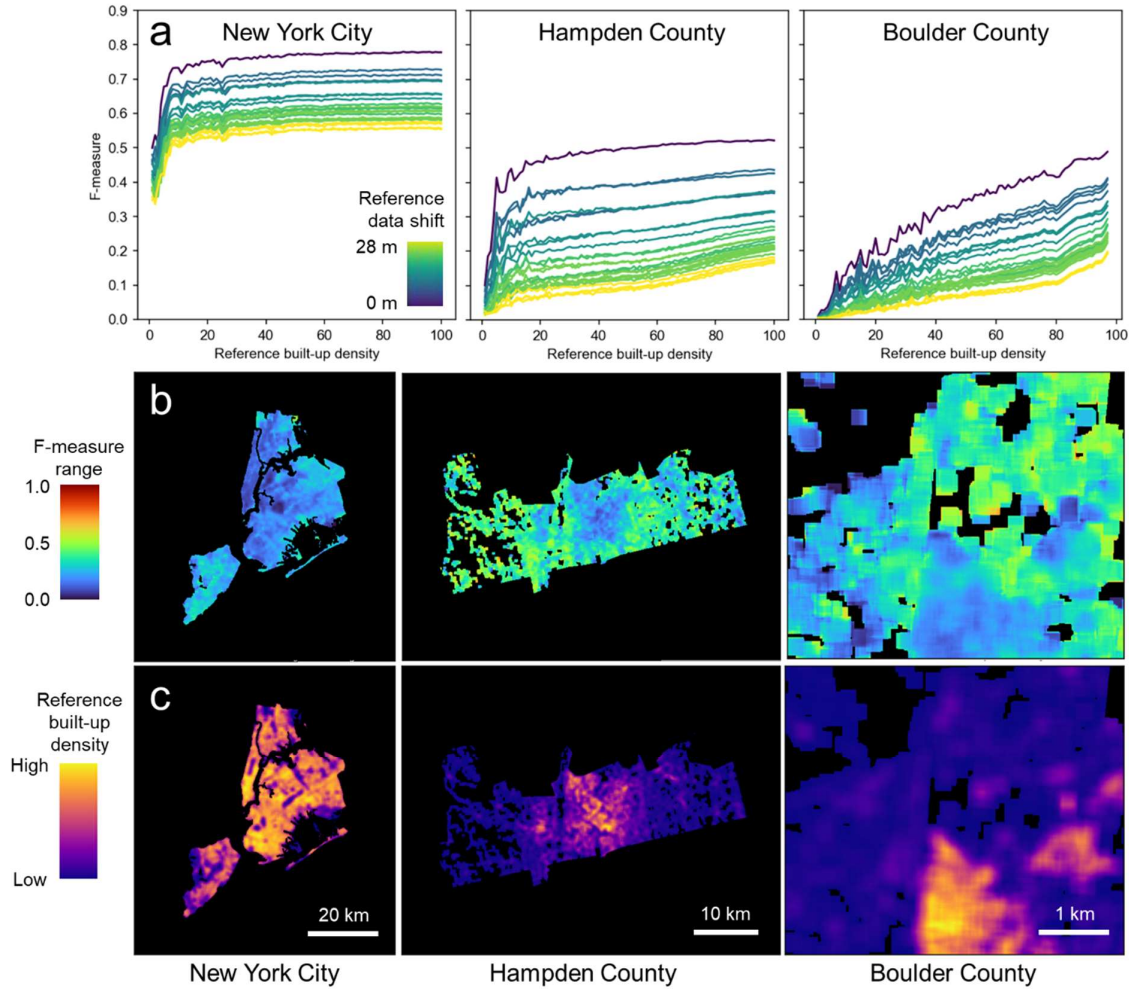

**Figure 11.** Sensitivity analysis of the focal accuracy surfaces to systematic offsets in GHS or reference data: (a) median trendlines of F-1 score per reference built-up area density stratum (using 100 equal-width bins), (b) maps of the range of focal F-1 scores per grid cell, based on the 25 grid shift combinations, and (c) the reference built-up surface density for comparison, illustrating that F-1 score ranges are lower in high-density regions. All data shown for New York City, Hampden County (Massachusetts), and for a subset of Boulder County (Colorado).

### 3.5.2. Sensitivity to the spatial support and to the analytical unit

Lastly, we tested the effect of increasing block size (i.e., analytical unit) and increasing spatial support (i.e., focal window size used as analysis extent) on our focal precision-recall signature plots (cf. Fig. 7). When increasing the block size, we observe increasing accuracy, in both precision and recall. This increase in the analytical unit is a mechanism to account for slight offsets in the gridded data, which would cause disagreement when calculated at the native resolution. We observe indeed increasing levels of both, precision and recall, in low-density but also high-density regions. Interestingly, increasing spatial support leaves overall precision and recall levels largely unaffected, but the dispersion of the focal accuracy metrics is reduced as we increase the spatial support, which is a general data aggregation effect. The same can be observed for the focal precision-recall signature plots calculated for a support of 5 km, for all 30 counties (Fig. A2), and jointly for increasing support and block size (Fig. A3). A similar effect is also notable when visualizing F-1 scores trajectories across the rural-urban continuum (Fig. A4): Extreme values disappear due to the increased aggregation of the focal accuracy estimates. Hence, it is recommended to keep the focal window size small – large enough to ensure a robust sample size, and small enough to capture the spatial details in accuracy variation, as a large focal window size will occlude fine spatial details in accuracy variation, and also increases the computational effort (i.e., more grid cells to take into account for each accuracy metric computation).

The increase of the analytical unit not only affects localized (focal) accuracy estimates, but also global (i.e., county-level accuracy estimates), exhibiting a similar trend, and potentially representing more realistic accuracy estimates (Table A3).

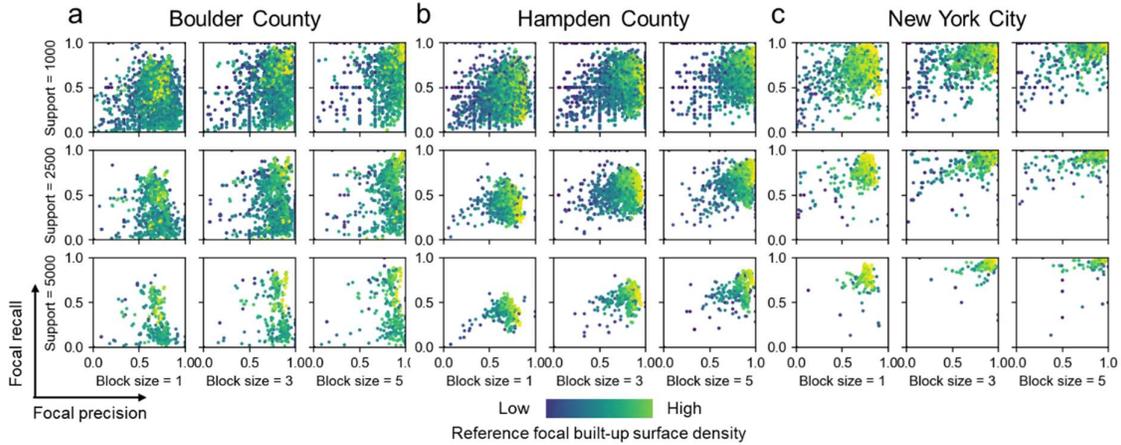

**Figure 12. Sensitivity to analytical scale and spatial support.** Shown are focal precision-recall signature plots, with varying analytical unit (i.e., block size from 1×1 grid cells, 3×3 grid cells, and 5×5 grid cells) in x-direction, and varying spatial support (i.e., focal window size, ranging from 1 km (top row) to 5 km (bottom row).

Finally, we also tested the effect of spatial support on the quantity agreement analysis (cf. Section 3.4). When comparing the regression analysis results of focal built-up density across different levels of spatial support, we observe increasing $R^2$ values as we increase the spatial support (an effect of extreme values being aggregated, see Fig. A5, and thus, reducing the impact of their residuals on the $R^2$), but we also observe fairly stable regression coefficients across different support levels (Table A4), indicating that the relationship between the reference and test built-up density is largely unaffected by the choice of the spatial support.

## 4. Conclusions

Herein, we presented a framework for the accuracy assessment of high resolution (i.e., 10 m) probabilistic built-up surface indicators. Specifically, we used machine learning-based probabilities of built-up area presence, as reported in the GHS-BUILT-S2 dataset and developed a multi-stage strategy for their evaluation against highly reliable building footprint data. The first stage is the analysis of the relationship between built-up probability and reference built-up fraction. While we found positive associations between these two variables, we also found that built-up probability cannot be directly translated into built-up area fractions, as indicated by the poor regression model performance (Section 3.1). In the second stage, we binarized both BUPROB and BUFRAC surfaces, and assessed for which threshold the agreement between these binarized surfaces maximized. We found that, for a threshold of 50%, applied to the GHS-BUILT-S2 data, the resulting binary built-up presence-absence layer shows the highest agreement with the reference dataset, which labeled any grid cell as built-up if it contains at least one building (or a part of it). This thresholding approach appeared to be an effective way to interpret and translate machine learning-based probabilities into a meaningful physical measure. In the third stage, we assessed the thematic and quantity agreement of the resulting built-up area layers and observed increasing accuracy from rural to urban settings, but we also observed much higher levels of accuracy (both precision and recall) in rural areas, when compared to accuracy assessments of earlier versions of the GHS-BUILT data.

While this multi-stage framework can be applied to similar datasets, we also proposed a novel visual-analytical tool for spatially explicit accuracy assessments of binary data, which are focal precision-recall signature plots (cf. Section 3.3). We also highlighted multiple methods for analyzing the sensitivity of accuracy estimates. Knowledge of the accuracy of the GHS-BUILT-S2 provides important guidance for an unbiased and informed interpretation of the dataset itself, and for follow-up data products such as the GHS-BUILT R2022A that are partially based on the GHS-BUILT-S2 (Schiavina et al. 2022).

A minor shortcoming of our analysis is the temporal gap between the MTBF-33 reference dataset (referred roughly to 2015) and the GHS-BUILT-S2 from 2018. The urban growth from 2015-2018, not measured by MTBF-33 would result in lower precision values. However, as we observe high precision levels throughout our analysis, we believe that this temporal discrepancy has only a minor effect on our findings. In future work, we will test different methods for the agreement maximization stage (e.g., ROC analysis; Green & Swets 1966), and apply this framework to larger spatial extents, within and outside of the United States.

## 5. Data and code availability

The GHS-BUILT-S2 R2020A dataset is available at http://dx.doi.org/10.2905/016D1A34-B184-42DC-B586-E10B915DD863. The MTBF-33 reference database is available at https://data.mendeley.com/datasets/w33vbvjtdy. Code for multi-resolution, global, zonal, and focal accuracy assessment is available at https://github.com/johannesuhl/local_accuracy.

## 6. Acknowledgments & Funding

Funding for this work was partially provided through the National Science Foundation (awards 1924670 and 2121976 to CU Boulder). This research benefited from support provided to the University of Colorado Population Center (CUPC, Project 2P2CHD066613-06) from the Eunice Kennedy Shriver Institute of Child Health Human and Human Development. The content is solely the responsibility of the authors and does not necessarily represent the official views of the National Institutes of Health or CUPC.

# 8. Appendix

**Table A1. Correlation coefficients between GHS built-up probability and built-up area fraction, as well as with focal accuracy / built-up surface density.**

| County | Pearson's correlation coefficient of built-up probability and | | | | |
|---|---|---|---|---|---|
| | BUFRAC | F1 | Precision | Recall | Reference built-up density |
| Anoka County | 0.284 | 0.086 | 0.042 | 0.084 | 0.521 |
| Baltimore County | 0.327 | 0.319 | 0.135 | 0.375 | 0.583 |
| Barnstable County | 0.407 | 0.168 | 0.114 | 0.119 | 0.464 |
| Berkshire County | 0.413 | 0.132 | 0.115 | 0.105 | 0.452 |
| Boulder County | 0.251 | 0.386 | 0.082 | 0.345 | 0.636 |
| Bristol County | 0.366 | 0.266 | 0.168 | 0.259 | 0.555 |
| Carver County | 0.290 | 0.047 | -0.097 | 0.211 | 0.468 |
| Dakota County | 0.287 | 0.267 | 0.203 | 0.234 | 0.518 |
| Dukes County | 0.406 | 0.164 | 0.127 | 0.126 | 0.496 |
| Essex County | 0.342 | 0.230 | 0.123 | 0.218 | 0.562 |
| Franklin County | 0.408 | 0.137 | 0.120 | 0.089 | 0.425 |
| Hampden County | 0.394 | 0.110 | 0.239 | 0.020 | 0.476 |
| Hampshire County | 0.398 | 0.134 | 0.159 | 0.074 | 0.427 |
| Hennepin County | 0.223 | 0.194 | 0.188 | 0.120 | 0.535 |
| Hillsborough County | 0.285 | 0.340 | 0.354 | 0.245 | 0.529 |
| Manatee County | 0.335 | 0.327 | 0.272 | 0.316 | 0.548 |
| Mecklenburg County | 0.328 | 0.173 | 0.119 | 0.142 | 0.291 |
| Middlesex County | 0.339 | 0.311 | 0.171 | 0.306 | 0.594 |
| Milwaukee County | 0.141 | 0.358 | 0.241 | 0.312 | 0.389 |
| Monmouth County | 0.376 | 0.208 | 0.169 | 0.164 | 0.509 |
| Nantucket County | 0.372 | 0.226 | 0.155 | 0.163 | 0.543 |
| Norfolk County | 0.340 | 0.228 | 0.125 | 0.219 | 0.503 |
| New York City | 0.331 | 0.442 | 0.334 | 0.428 | 0.643 |
| Plymouth County | 0.384 | 0.204 | 0.132 | 0.187 | 0.482 |
| Ramsey County | 0.282 | 0.207 | 0.201 | 0.117 | 0.410 |
| Sarasota County | 0.322 | 0.260 | 0.229 | 0.240 | 0.503 |
| Suffolk County | 0.246 | 0.340 | 0.254 | 0.347 | 0.526 |
| Vanderburgh County | 0.370 | 0.256 | 0.164 | 0.204 | 0.555 |
| Washington County | 0.288 | 0.174 | 0.058 | 0.238 | 0.519 |
| Worcester County | 0.367 | 0.185 | 0.140 | 0.146 | 0.461 |

**Table A2. Fitting polynomial functions to estimate built-up fraction from built-up probabilities, per county, and for polynomials of degree 1 to 4.**

| Polynomial degree | 1st | | 2nd | | 3rd | | 4th | |
|---|---|---|---|---|---|---|---|---|
| County | RMSE | $R^2$ | RMSE | $R^2$ | RMSE | R2 | RMSE | $R^2$ |
| Anoka County | 29.136 | 0.081 | 28.588 | 0.115 | 28.307 | 0.132 | 28.268 | 0.135 |
| Baltimore County | 29.752 | 0.107 | 29.090 | 0.147 | 28.792 | 0.164 | 28.783 | 0.164 |
| Barnstable County | 24.492 | 0.166 | 24.140 | 0.190 | 24.072 | 0.194 | 24.050 | 0.196 |
| Berkshire County | 26.365 | 0.170 | 26.041 | 0.191 | 25.997 | 0.193 | 25.997 | 0.193 |
| Boulder County | 30.367 | 0.063 | 30.100 | 0.080 | 30.005 | 0.085 | 29.998 | 0.086 |
| Bristol County | 27.898 | 0.134 | 27.255 | 0.174 | 27.198 | 0.177 | 27.185 | 0.178 |
| Carver County | 29.310 | 0.084 | 28.729 | 0.120 | 28.407 | 0.140 | 28.368 | 0.142 |
| Dakota County | 30.341 | 0.082 | 29.908 | 0.108 | 29.570 | 0.128 | 29.544 | 0.130 |
| Dukes County | 23.454 | 0.165 | 23.168 | 0.185 | 23.112 | 0.189 | 23.100 | 0.190 |
| Essex County | 27.865 | 0.117 | 27.593 | 0.134 | 27.557 | 0.136 | 27.545 | 0.137 |
| Franklin County | 25.984 | 0.167 | 25.611 | 0.190 | 25.561 | 0.193 | 25.559 | 0.194 |
| Hampden County | 27.593 | 0.155 | 27.183 | 0.180 | 27.116 | 0.184 | 27.114 | 0.184 |
| Hampshire County | 26.714 | 0.158 | 26.397 | 0.178 | 26.339 | 0.182 | 26.337 | 0.182 |
| Hennepin County | 30.948 | 0.050 | 30.582 | 0.072 | 30.438 | 0.081 | 30.421 | 0.082 |
| Hillsborough County | 29.727 | 0.081 | 29.508 | 0.095 | 29.497 | 0.095 | 29.483 | 0.096 |
| Manatee County | 29.343 | 0.112 | 29.195 | 0.121 | 29.182 | 0.122 | 29.164 | 0.123 |
| Mecklenburg County | 30.514 | 0.108 | 30.038 | 0.135 | 29.791 | 0.150 | 29.779 | 0.150 |
| Middlesex County | 28.514 | 0.115 | 28.205 | 0.134 | 28.181 | 0.135 | 28.173 | 0.136 |
| Milwaukee County | 31.189 | 0.020 | 30.589 | 0.057 | 30.580 | 0.058 | 30.501 | 0.063 |
| Monmouth County | 27.446 | 0.142 | 26.481 | 0.201 | 26.375 | 0.207 | 26.327 | 0.210 |
| Nantucket County | 24.205 | 0.138 | 23.958 | 0.156 | 23.930 | 0.158 | 23.927 | 0.158 |
| Norfolk County | 28.612 | 0.116 | 28.183 | 0.142 | 28.139 | 0.145 | 28.122 | 0.146 |
| New York City | 30.049 | 0.110 | 29.095 | 0.165 | 29.015 | 0.170 | 28.948 | 0.174 |
| Plymouth County | 26.442 | 0.148 | 25.938 | 0.180 | 25.863 | 0.185 | 25.844 | 0.186 |
| Ramsey County | 29.582 | 0.080 | 29.317 | 0.096 | 29.204 | 0.103 | 29.184 | 0.104 |
| Sarasota County | 29.233 | 0.104 | 29.107 | 0.112 | 29.088 | 0.113 | 29.077 | 0.113 |
| Suffolk County | 30.991 | 0.061 | 30.506 | 0.090 | 30.493 | 0.091 | 30.466 | 0.092 |
| Vanderburgh County | 28.985 | 0.137 | 28.545 | 0.163 | 28.471 | 0.167 | 28.460 | 0.168 |
| Washington County | 29.124 | 0.083 | 28.563 | 0.118 | 28.254 | 0.137 | 28.213 | 0.139 |
| Worcester County | 27.916 | 0.135 | 27.609 | 0.154 | 27.578 | 0.156 | 27.573 | 0.156 |

Table A3. Sensitivity of overall acc met to analytical unit – block sizes

| County | Precision (1x1) | Precision (3x3) | Precision (5x5) | Recall (1x1) | Recall (3x3) | Recall (5x5) | F1 (1x1) | F1 (3x3) | F1 (5x5) |
|---|---|---|---|---|---|---|---|---|---|
| Anoka County | 0,612 | 0,764 | 0,841 | 0,554 | 0,673 | 0,727 | 0,581 | 0,715 | 0,780 |
| Baltimore County | 0,672 | 0,816 | 0,858 | 0,669 | 0,792 | 0,842 | 0,671 | 0,804 | 0,850 |
| Barnstable County | 0,666 | 0,835 | 0,898 | 0,567 | 0,775 | 0,872 | 0,613 | 0,803 | 0,884 |
| Berkshire County | 0,731 | 0,845 | 0,882 | 0,461 | 0,606 | 0,692 | 0,565 | 0,706 | 0,776 |
| Boulder County | 0,683 | 0,819 | 0,862 | 0,320 | 0,438 | 0,478 | 0,436 | 0,571 | 0,615 |
| Bristol County | 0,700 | 0,844 | 0,889 | 0,624 | 0,790 | 0,860 | 0,660 | 0,816 | 0,874 |
| Carver County | 0,610 | 0,758 | 0,820 | 0,568 | 0,657 | 0,693 | 0,588 | 0,704 | 0,752 |
| Dakota County | 0,626 | 0,769 | 0,830 | 0,716 | 0,858 | 0,909 | 0,668 | 0,811 | 0,867 |
| Dukes County | 0,604 | 0,746 | 0,807 | 0,560 | 0,749 | 0,826 | 0,581 | 0,748 | 0,817 |
| Essex County | 0,714 | 0,879 | 0,920 | 0,554 | 0,770 | 0,857 | 0,624 | 0,821 | 0,887 |
| Franklin County | 0,676 | 0,772 | 0,810 | 0,506 | 0,648 | 0,729 | 0,579 | 0,705 | 0,768 |
| Hampden County | 0,747 | 0,868 | 0,912 | 0,394 | 0,583 | 0,728 | 0,516 | 0,698 | 0,809 |
| Hampshire County | 0,719 | 0,832 | 0,872 | 0,447 | 0,607 | 0,711 | 0,552 | 0,702 | 0,784 |
| Hennepin County | 0,645 | 0,795 | 0,862 | 0,606 | 0,759 | 0,805 | 0,625 | 0,777 | 0,832 |
| Hillsborough County | 0,583 | 0,735 | 0,778 | 0,671 | 0,842 | 0,907 | 0,624 | 0,785 | 0,838 |
| Manatee County | 0,658 | 0,799 | 0,833 | 0,721 | 0,858 | 0,900 | 0,688 | 0,828 | 0,865 |
| Mecklenburg County | 0,763 | 0,873 | 0,916 | 0,516 | 0,663 | 0,778 | 0,615 | 0,753 | 0,841 |
| Middlesex County | 0,717 | 0,870 | 0,914 | 0,287 | 0,386 | 0,430 | 0,410 | 0,535 | 0,585 |
| Milwaukee County | 0,697 | 0,881 | 0,925 | 0,289 | 0,437 | 0,483 | 0,408 | 0,584 | 0,635 |
| Monmouth County | 0,651 | 0,792 | 0,848 | 0,414 | 0,495 | 0,528 | 0,507 | 0,609 | 0,651 |
| Nantucket County | 0,585 | 0,733 | 0,795 | 0,586 | 0,800 | 0,874 | 0,585 | 0,765 | 0,832 |
| Norfolk County | 0,712 | 0,868 | 0,915 | 0,585 | 0,784 | 0,875 | 0,642 | 0,824 | 0,895 |
| New York City | 0,771 | 0,916 | 0,929 | 0,824 | 0,944 | 0,975 | 0,797 | 0,929 | 0,951 |
| Plymouth County | 0,681 | 0,827 | 0,881 | 0,571 | 0,750 | 0,837 | 0,621 | 0,787 | 0,859 |
| Ramsey County | 0,682 | 0,828 | 0,886 | 0,596 | 0,793 | 0,859 | 0,636 | 0,810 | 0,872 |
| Sarasota County | 0,663 | 0,823 | 0,872 | 0,710 | 0,865 | 0,922 | 0,685 | 0,844 | 0,896 |
| Suffolk County | 0,742 | 0,899 | 0,914 | 0,720 | 0,914 | 0,955 | 0,731 | 0,906 | 0,934 |
| Vanderburgh County | 0,738 | 0,862 | 0,901 | 0,589 | 0,754 | 0,832 | 0,655 | 0,804 | 0,865 |
| Washington County | 0,621 | 0,762 | 0,823 | 0,588 | 0,691 | 0,734 | 0,604 | 0,725 | 0,776 |
| Worcester County | 0,727 | 0,854 | 0,896 | 0,227 | 0,327 | 0,385 | 0,346 | 0,473 | 0,538 |

Table A4. Regression analysis of focal reference and GHS built-up quantity per county and support level.

| Spatial support | 1 km | | | 2.5 km | | | 5 km | | |
|---|---|---|---|---|---|---|---|---|---|
| County | Slope | Intercept | $R^2$ | Slope | Intercept | $R^2$ | Slope | Intercept | $R^2$ |
| All counties | 0.874 | 0.203 | 0.868 | 0.878 | 0.108 | 0.883 | 0.886 | 0.033 | 0.888 |
| Anoka County | 1.102 | 0.456 | 0.946 | 1.111 | 0.390 | 0.971 | 1.122 | 0.321 | 0.980 |
| Baltimore County | 1.020 | -0.155 | 0.953 | 1.050 | -0.369 | 0.979 | 1.061 | -0.426 | 0.990 |
| Barnstable County | 0.847 | 0.007 | 0.928 | 0.864 | -0.078 | 0.951 | 0.888 | -0.217 | 0.959 |
| Berkshire County | 0.602 | 0.064 | 0.928 | 0.614 | 0.026 | 0.971 | 0.629 | 0.004 | 0.985 |
| Boulder County | 0.655 | -0.104 | 0.875 | 0.585 | 0.059 | 0.821 | 0.481 | 0.151 | 0.742 |
| Bristol County | 0.936 | -0.314 | 0.966 | 0.954 | -0.400 | 0.977 | 0.961 | -0.420 | 0.981 |
| Carver County | 1.177 | 0.047 | 0.914 | 1.220 | -0.058 | 0.962 | 1.251 | -0.116 | 0.984 |
| Dakota County | 1.078 | 0.410 | 0.948 | 1.101 | 0.258 | 0.976 | 1.116 | 0.179 | 0.985 |
| Dukes County | 0.864 | 0.184 | 0.964 | 0.878 | 0.130 | 0.983 | 0.889 | 0.095 | 0.991 |
| Essex County | 0.810 | -0.335 | 0.967 | 0.821 | -0.397 | 0.981 | 0.826 | -0.418 | 0.987 |
| Franklin County | 0.702 | 0.075 | 0.958 | 0.726 | 0.029 | 0.978 | 0.753 | -0.001 | 0.983 |
| Hampden County | 0.441 | 0.395 | 0.858 | 0.435 | 0.515 | 0.863 | 0.445 | 0.440 | 0.902 |
| Hampshire County | 0.570 | 0.161 | 0.908 | 0.590 | 0.088 | 0.950 | 0.611 | 0.031 | 0.964 |
| Hennepin County | 0.983 | 1.470 | 0.938 | 0.990 | 1.349 | 0.957 | 0.991 | 1.258 | 0.972 |
| Hillsborough County | 1.011 | 1.454 | 0.932 | 1.037 | 0.991 | 0.968 | 1.047 | 0.809 | 0.980 |
| Manatee County | 1.046 | 0.434 | 0.965 | 1.048 | 0.236 | 0.985 | 1.051 | 0.187 | 0.991 |
| Mecklenburg County | 0.549 | 1.913 | 0.686 | 0.520 | 2.281 | 0.746 | 0.490 | 2.608 | 0.792 |
| Middlesex County | 0.873 | -0.798 | 0.958 | 0.891 | -0.938 | 0.971 | 0.887 | -0.896 | 0.978 |
| Milwaukee County | 0.654 | 1.462 | 0.697 | 0.693 | 0.184 | 0.785 | 0.715 | -0.518 | 0.856 |
| Monmouth County | 0.994 | 0.939 | 0.930 | 1.033 | 0.565 | 0.967 | 1.045 | 0.413 | 0.982 |
| Nantucket County | 0.904 | 0.412 | 0.958 | 0.936 | 0.196 | 0.980 | 0.964 | 0.102 | 0.989 |
| Norfolk County | 0.853 | -0.322 | 0.951 | 0.877 | -0.539 | 0.968 | 0.914 | -0.846 | 0.976 |
| New York City | 1.054 | 0.601 | 0.900 | 1.071 | 0.008 | 0.914 | 1.099 | -0.876 | 0.929 |
| Plymouth County | 0.823 | 0.065 | 0.946 | 0.831 | 0.017 | 0.973 | 0.831 | 0.001 | 0.982 |
| Ramsey County | 0.811 | 2.616 | 0.918 | 0.812 | 2.460 | 0.948 | 0.829 | 1.992 | 0.963 |
| Sarasota County | 1.018 | 0.689 | 0.963 | 1.023 | 0.484 | 0.986 | 1.030 | 0.343 | 0.994 |
| Suffolk County | 0.947 | 0.818 | 0.938 | 0.946 | 0.762 | 0.950 | 0.946 | 0.690 | 0.955 |
| Vanderburgh County | 0.760 | 0.342 | 0.947 | 0.756 | 0.336 | 0.969 | 0.756 | 0.306 | 0.979 |
| Washington County | 1.115 | 0.131 | 0.928 | 1.140 | 0.041 | 0.957 | 1.177 | -0.081 | 0.971 |
| Worcester County | 0.598 | 0.010 | 0.900 | 0.601 | -0.057 | 0.917 | 0.616 | -0.189 | 0.914 |

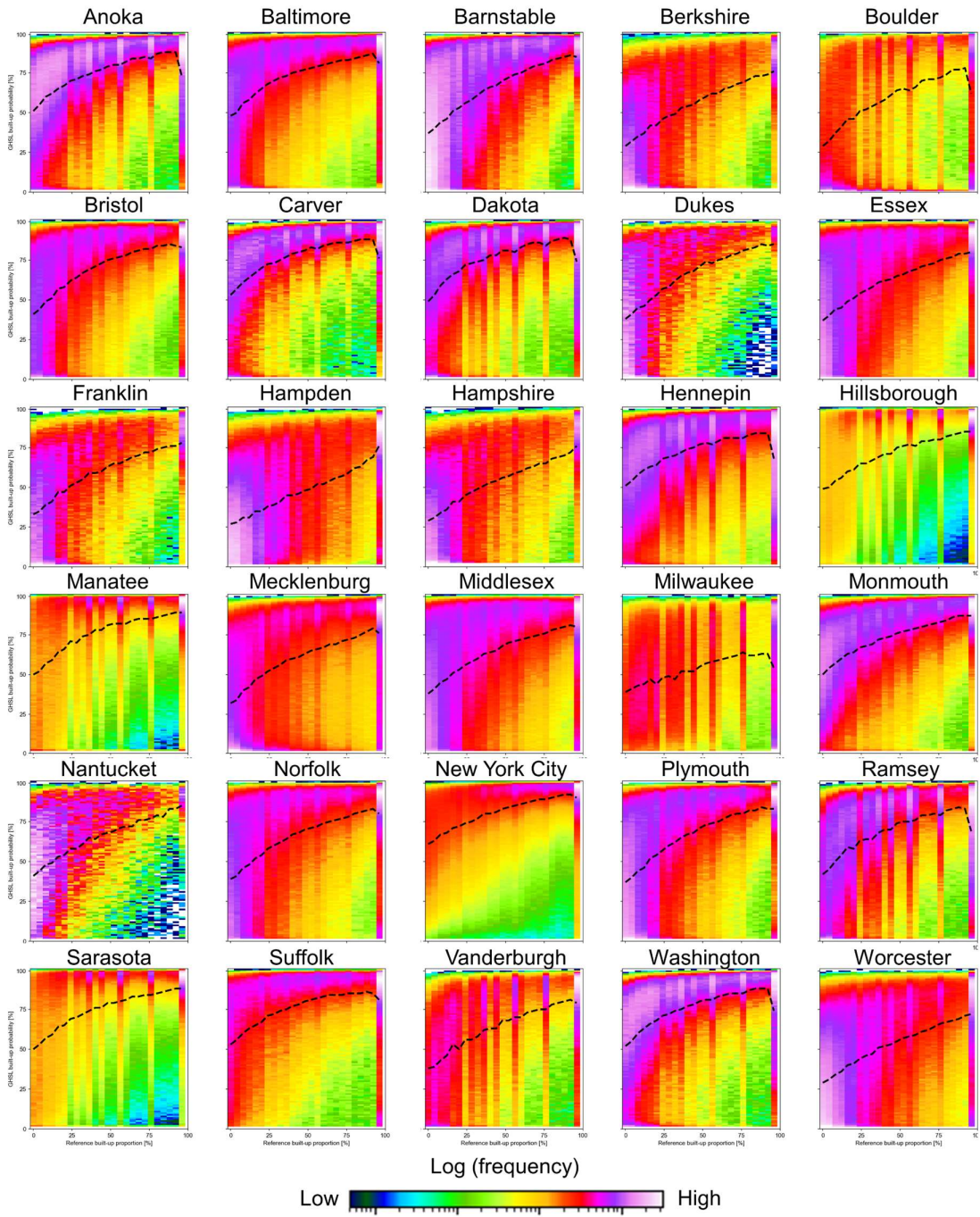

**Figure A1. Bi-variate frequency distributions of built-up fraction (x-axis) and built-up probability (y-axis) for all 10 m grid cells within each of the 30 counties under study.**

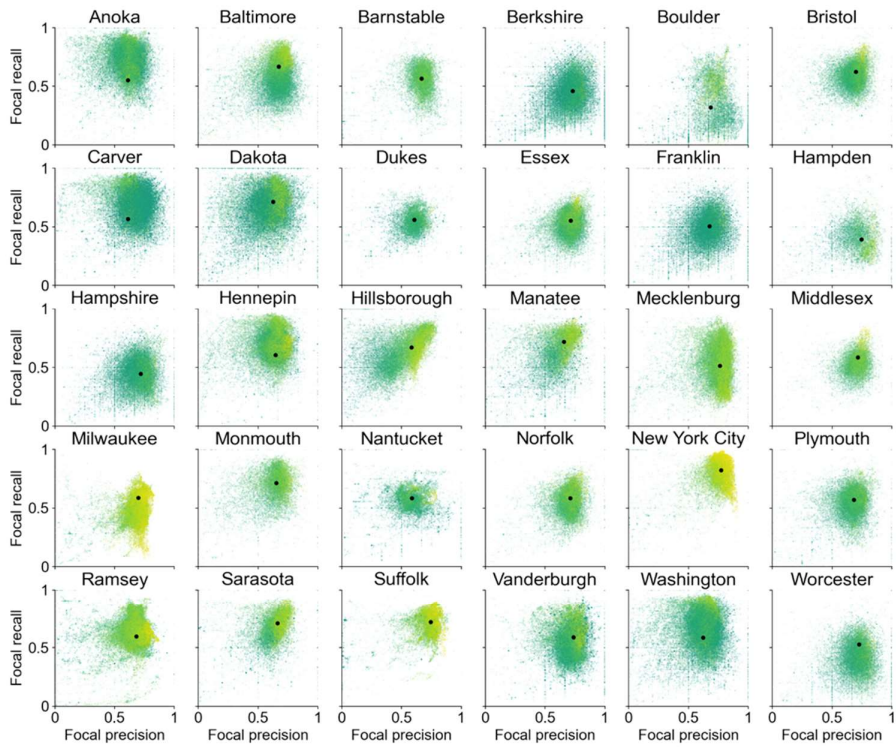
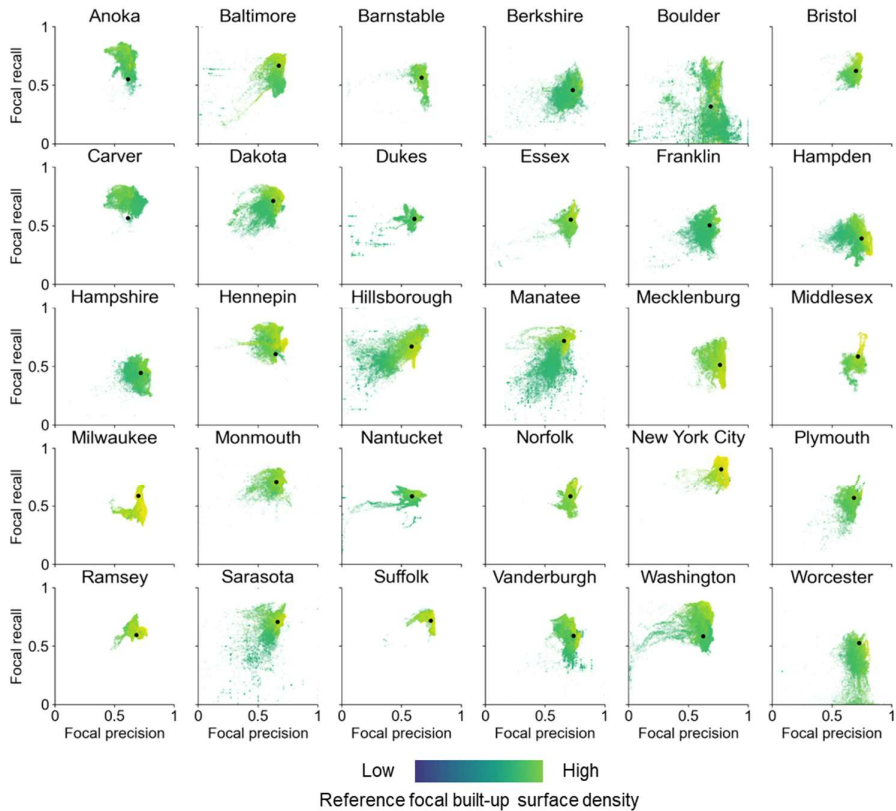
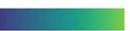

**Figure A2: Focal precision-recall signature plots for a spatial support of 1 km and 5 km.**

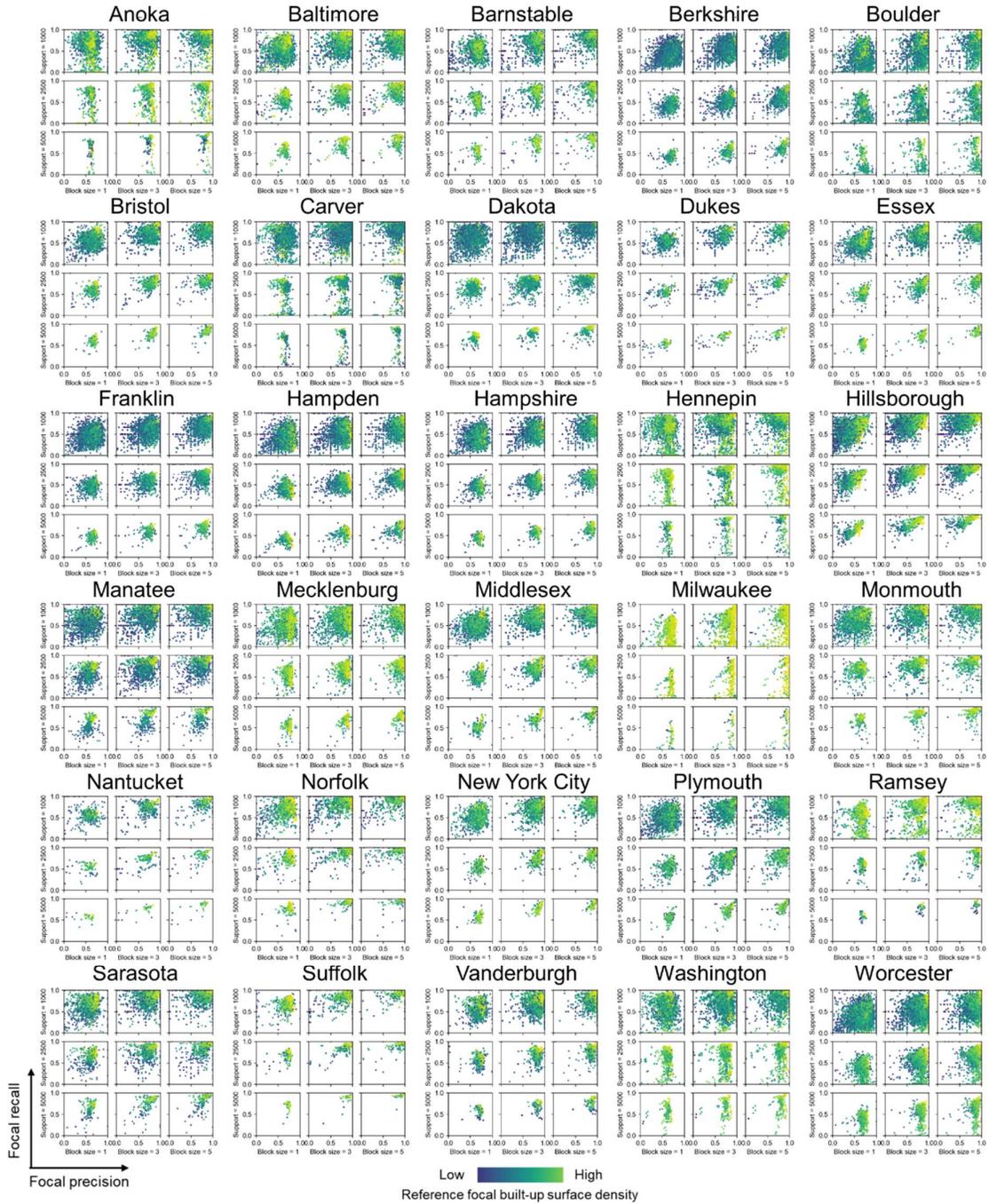

**Figure A3. Scale sensitivity analyses per county.** Shown are focal precision-recall signature plots, with varying analytical unit (i.e., block size from 1×1 grid cells, 3×3 grid cells, and 5×5 grid cells) in x-direction, and varying spatial support (i.e., focal window size, ranging from 1 km (top row) to 5 km (bottom row).

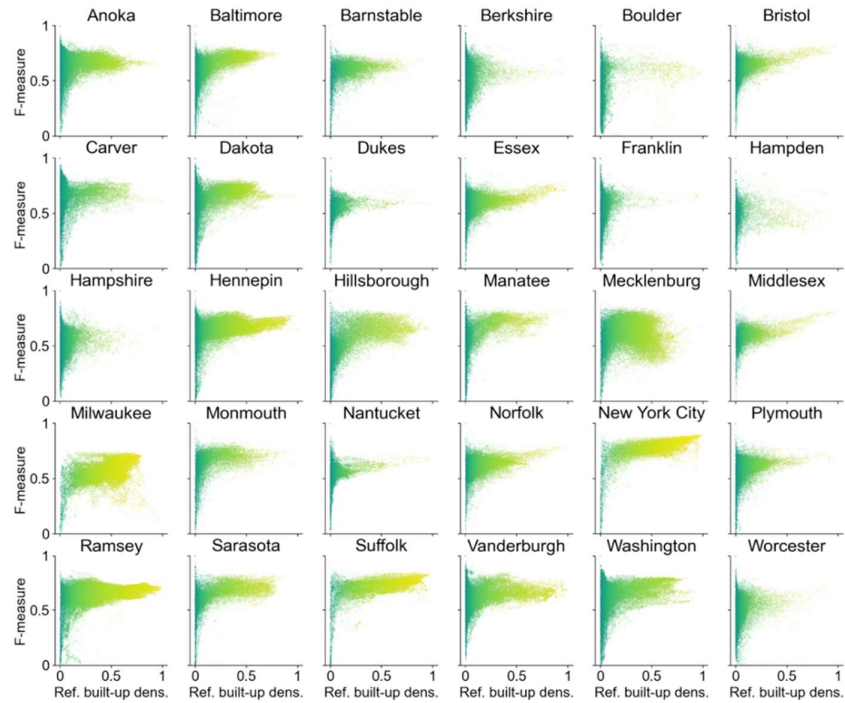
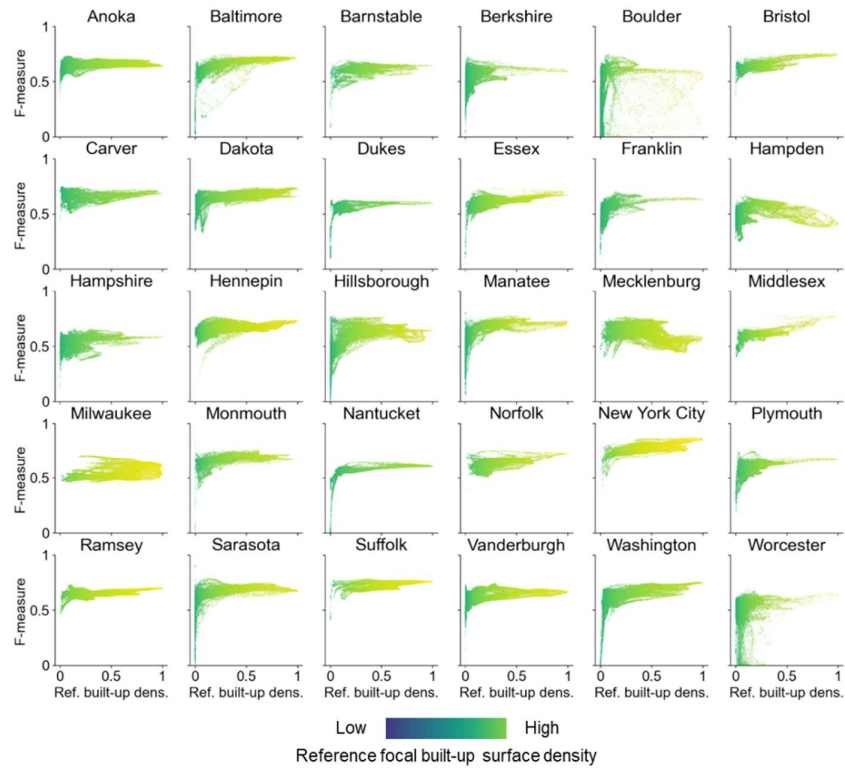

**Figure A4: Focal built-up density vs. F-measure plots for a spatial support of 1 km and 5 km.**

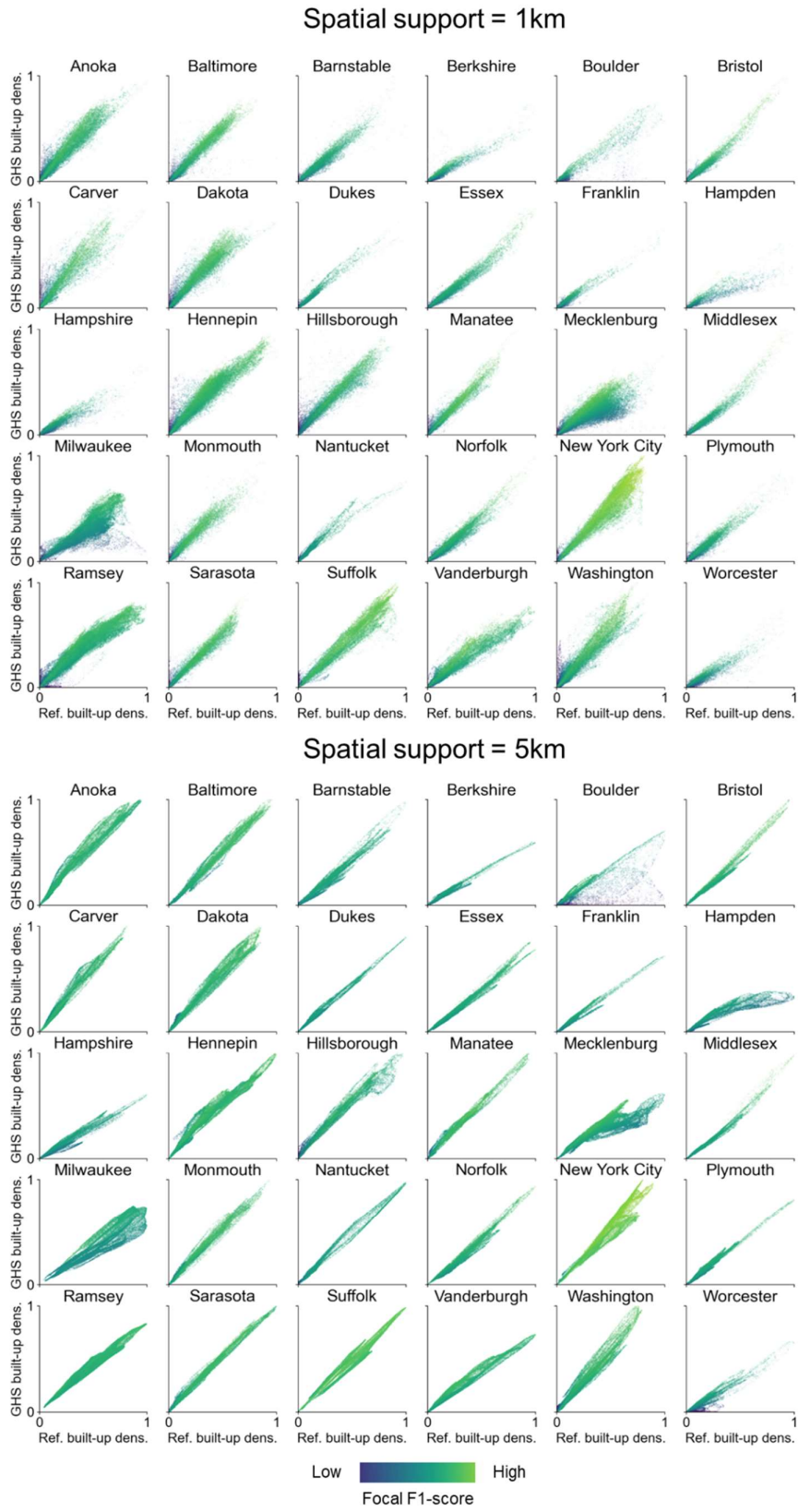

**Figure A5: Focal reference built-up quantity vs. GHS built-up quantity per county, for a spatial support of 1 km and 5 km.**